# Vectorial adaptive optics

Chao He[†*], Jacopo Antonello[†*] and Martin J. Booth[*]

*Department of Engineering Science, University of Oxford, Parks Road, Oxford, OX1 3PJ, UK*

[†]*These authors contributed equally to this work*

[*]*Corresponding authors:* chao.he@eng.ox.ac.uk; jacopo.antonello@eng.ox.ac.uk; martin.booth@eng.ox.ac.uk

**Adaptive optics normally concerns the feedback correction of phase aberrations. Such correction has been of benefit in various optical systems, with applications ranging in scale from astronomical telescopes to super-resolution microscopes. Here we extend this powerful tool into the vectorial domain, encompassing higher-dimensional feedback correction of both polarisation and phase. This technique is termed vectorial adaptive optics (V-AO). We show that V-AO can be implemented using sensor feedback, indirectly using sensorless AO, or in hybrid form combining aspects of both. We validate improvements in both vector field state and the focal quality of an optical system, through correction for commonplace vectorial aberration sources, ranging from objective lenses to biological samples. This technique pushes the boundaries of traditional scalar beam shaping by providing feedback control of extra vectorial degrees of freedom. This paves the way for next generation AO functionality by manipulating the complex vectorial field.**

**Introduction**

Phase aberrations affect the performance of many optical systems. Adaptive optics (AO) is widely used to perform feedback correction of these aberrations in a range of applications, from the inter-galactic scale of astronomical telescopes [1] to the molecular level in super-resolution microscopy [2, 3]. However, in many systems, polarisation aberrations play an even more crucial role. These aberrations lead to polarisation errors and extra (dynamic and geometric) phase distortion (see **Supplementary Note 1**) that can be introduced, for example, when focusing through stressed optical elements, due to Fresnel's effects or induced via polarising effects in materials or biological tissues [4-7]. These effects directly alter the state of polarisation (SOP) of the light field and the focal quality, hence affecting vectorial information analysis and degrading the system resolution in ways that compound the effects of phase aberrations. Considered jointly as "vectorial aberrations" (see Fig. 1a), these polarisation and phase errors limit the performance of many vector sensitive or high-resolution optical systems.

Incorrect vector states in the illumination or detection beams are greatly detrimental for polarisation sensitive microscopes, including Stokes/Mueller confocal microscopes [7], second/third harmonic generation microscopes [8] and super-resolution fluorescence polarisation microscopy [9]. Such effects are vital, for example, to provide correct vectorial information in label-free cancer detection using Stokes/Mueller microscopy. Incorrect polarisation states also disrupt the interference at the focus, hence affecting imaging resolution. This is particularly important in sensitive super-resolution methods, for example, in the creation of the zero intensity centre of the ring-shaped STED microscopy or MINFLUX beams [10, 11, 12]; in the interference light fields of the SIM or 4Pi microscopes [13, 14]; and can also affect performance of other common microscopes [15]. Such effects would be exacerbated in deep sample imaging where there are compounded polarisation and phase errors.

As these vectorial aberrations vary with the imaging scenario, field positions, and specimen type, adaptive feedback correction is essential to obtain optimal system performance. We therefore need to extend the concepts of conventional AO to the joint compensation of polarisation and phase aberrations through introduction of the concept of vectorial adaptive optics (V-AO). Conventional phase AO requires a method of phase measurement – either through a wavefront sensor or indirect optimisation methods ("sensorless AO" for short) [2, 3] – to determine the input aberration and a method of phase compensation; whereas V-AO requires the sensing and correction of the vectorial aberration. There are significant challenges in extending existing AO methods to this higher-dimensional analogue to conventional phase correction.

In order to meet these challenges, we have implemented and validated V-AO correction through three methods: A) sensor-based, B) quasi-sensorless, C) modal-sensorless; conceptual sketches of the feedback process are shown in Fig. 1b. We describe in detail the three methods and their properties, with demonstrations of the improvement of both vector field and focus after correction of commonplace vectorial aberrations. Our results indicate that V-AO can suppress vectorial aberrations, thus enhancing the toolbox for applications beyond those of traditional AO.

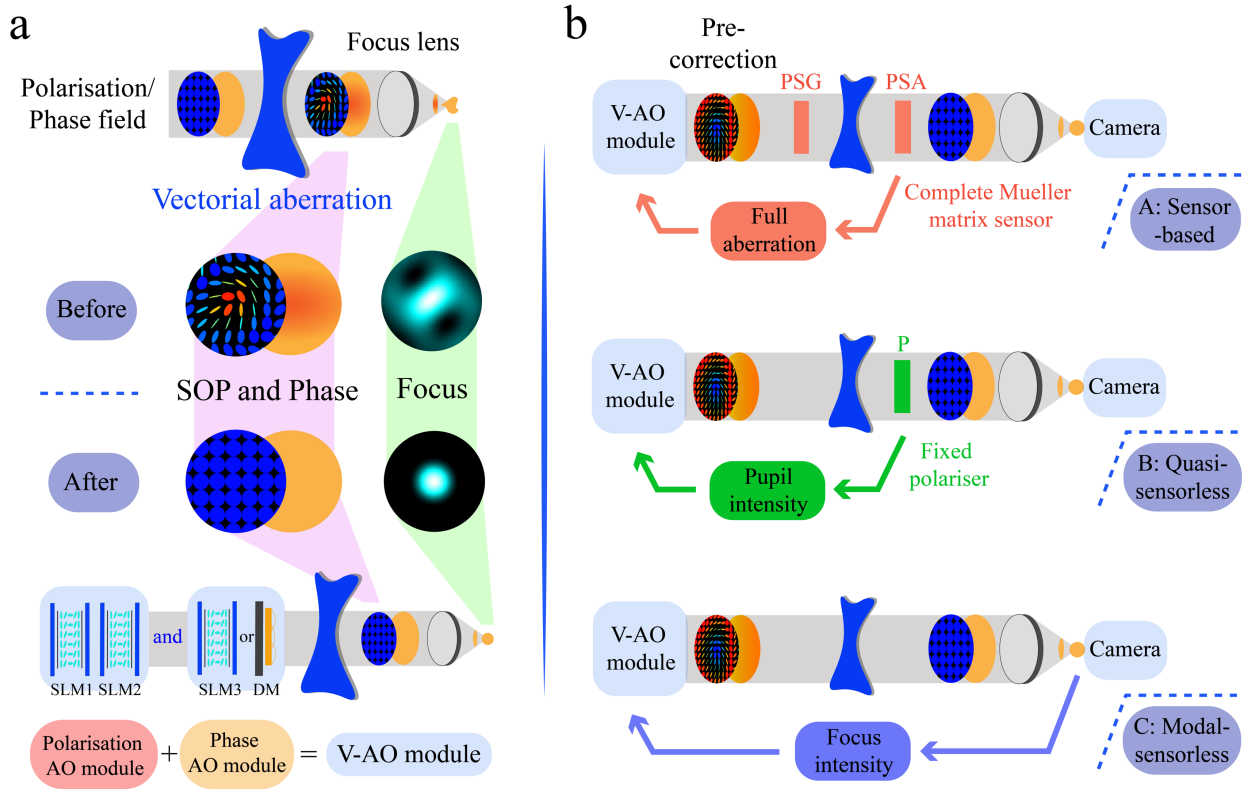

**Figure 1: Vectorial aberrations, vectorial adaptive optics (V-AO), and correction methods ranging from sensor-based to sensorless.** (a) Schematic of a system affected by vectorial aberration and a V-AO compensated system. The V-AO module encompasses a polarisation module and a phase module. Here the V-AO module pre-corrects the vectorial aberration. The state of polarisation (SOP) and phase profiles, as well as focus profile are given as sketches illustrating states 'before' and 'after' V-AO correction. (b) Sketches of three different V-AO correction methods. Method A (sensor-based V-AO): A MM sensor is used to measure the aberration then give feedback to V-AO module for pre-correction, hence leading to the desired light field output. PSG: polarisation state generator; PSA: polarisation state analyser. Method B (quasi-sensorless V-AO): A polariser is used to filter the intensity before the pupil, and the pupil intensity profile after the polariser serves as feedback to manipulate the V-AO module to correct the aberration. P: Polariser. Method C (modal-sensorless V-AO): Focal intensity is used as proxy for optimising the output phase and SOPs of the beam hence completing the V-AO loop.

## Results

In contrast to conventional AO, V-AO requires a method to determine the full vectorial properties, encompassing polarisation and phase, and a mechanism for its control. Conventional Mueller matrix (MM) polarimetry plays a role here, as the "sensor" for determination of the MM (which is equivalent to the polarisation aberration) as well as the output state of polarisation (SOP). Sensorless V-AO methods can also be employed, allowing for example model-based optimisation of focal quality through adjustment of polarisation and phase states. Based on these principles, we introduce a suite of V-AO techniques that can be applied in different situations. Central to this is a versatile vectorial field compensation system consisting of two liquid crystal spatial light modulators (SLMs) to fully control the output SOP [16, 17] and a deformable mirror (DM) to compensate for phase (Fig. 1a) [2, 3]. Similar multi adaptive element systems have been used for the purpose of complex beam generation [18, 19, 20], whereas here we emphasise the use of such manipulators as the aberration corrector in a feedback AO system.

**Sensor-based vectorial adaptive optics (Method A)**

At the heart of the sensor-based V-AO implementation is an imaging MM polarimeter, which is able to extract the full polarisation properties across the profile of a beam or an object [7] (Fig. 1b). A complete MM polarimeter consists of a polarisation state generator (PSG) and a polarisation state analyser (PSA) [7]. Following the MM measurement of the object using the PSG and PSA, the SLMs are set such that the output SOP after propagation through the object is spatially uniform. This is achieved by setting the patterns displayed by precisely calibrated SLMs (see **Supplementary Note 2**), so that the SLMs introduce a pre-compensation aberration, leading to uniform output SOP. Phase aberration can also be introduced by the object, while additionally further phase variations are introduced by the SLMs themselves [21] (see **Supplementary Note 1**). Full correction is achieved by first correcting the SOP aberration using the SLMs and then applying phase only correction via conventional sensorless AO.

To illustrate some peculiarities of full vectorial aberrations (comprising both SOP and phase) we first introduce a vectorial aberration using the V-AO module. Three notable cases are depicted in Fig. 2a: 1) a beam with a uniform SOP was perturbed by such an aberration (V-AO off), resulting in both a disordered SOP in the pupil as well as a distorted focus. 2) (Phase AO on) we applied phase-only sensorless AO with the DM which results again in an aberrated focus. This verifies that traditional phase-only AO cannot fully compensate a full vectorial aberration, because a disordered polarisation state affects the constructive interference required for perfect focussing [22]. 3) Finally, the initial vectorial aberration introduced by the V-AO module is removed as following the correction procedure, resulting in a spatially near-uniform SOP in the pupil and a diffraction limited focus (V-AO on). We illustrate in Fig. 2b on the Poincaré sphere the distribution of the SOP in along a line in the cross-section of the pupil as well as cross-sections of the foci in Fig. 2a. Corresponding theoretical validation and simulations can be found in **Supplementary Note 1**. The well-matched results (for simulation and experiments) also validated the capabilities of V-AO approach.

We then chose a graded index (GRIN) lens for demonstration of a real vectorial aberration. GRIN lenses are widely used for compact imaging systems and microscopy; applications span across connectors for quantum chips to biopsy probes for clinical diagnosis. By nature of their manufacture, GRIN lenses suffer from a rotationally symmetric birefringence variation that is concomitant with their symmetrical graded index profile [4]. This property is considered as a nuisance as it introduces a vectorial perturbation that disrupts GRIN lens based imaging systems [4]; these perturbations cannot be compensated via traditional phase AO. Therefore, widespread adoption of GRIN optics is hindered in sensitive systems, such as for compact super-resolution or polarisation contrast imaging systems [4]. Similar experiments are conducted following the same process in Fig. 2a. Fig. 2c shows results comparing before (V-AO off) and after correction (V-AO on). The vector fields, the correction patterns on the DM, as well as focal spot comparisons are given. More detailed analysis can be found in **Supplementary Note 3**.

These demonstrations show that the feedback sensor-based V-AO method (method A) can improve the performance of an optical system both in terms of uniformity of the SOP in the pupil and distortions at the focus.

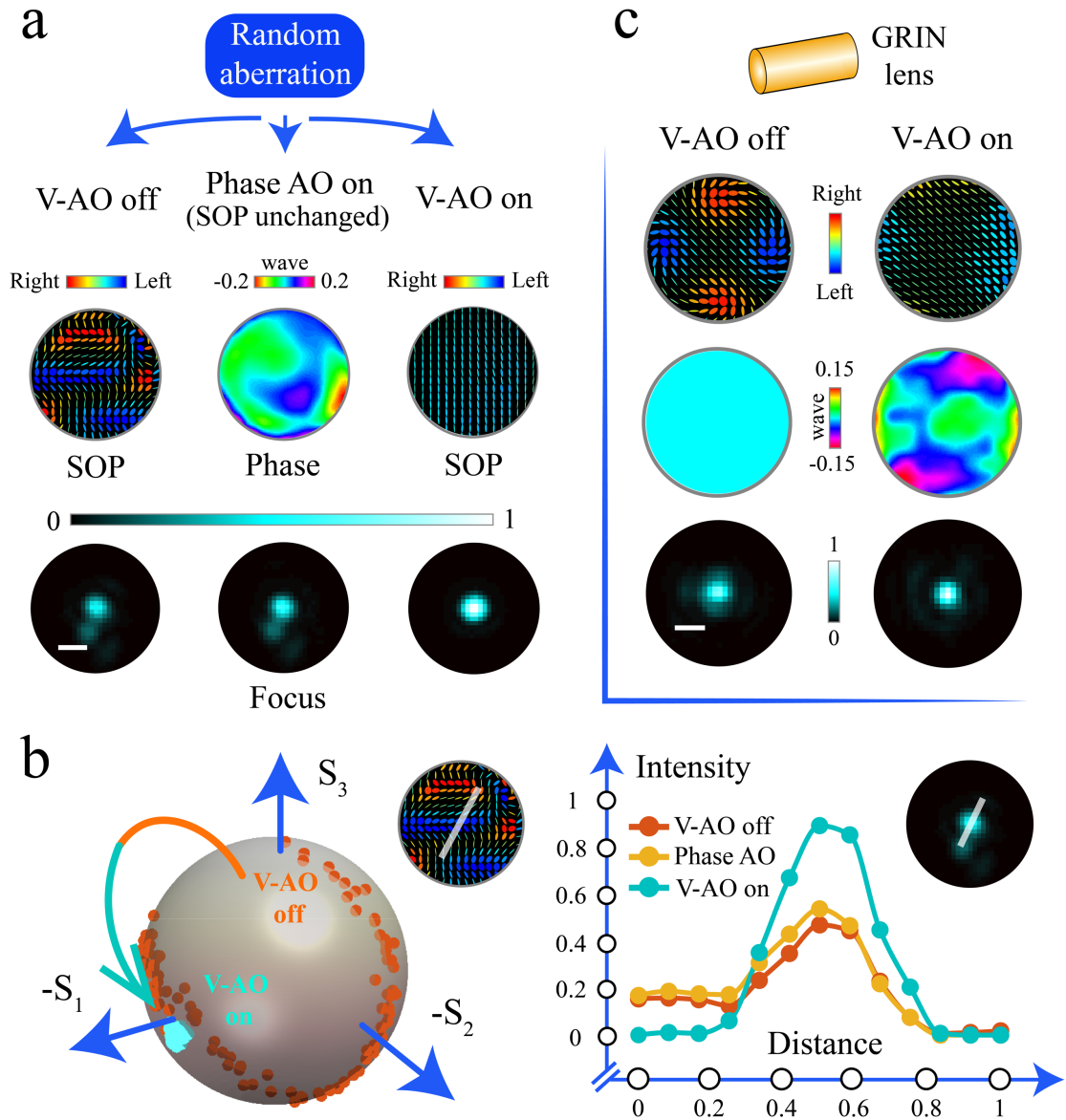

**Figure 2: Method A – sensor-based vectorial adaptive optics.** (a) Random aberration case, showing V-AO off, phase AO only, and V-AO on. The vector field SOP profile, phase on the SM, profile of the foci are given. (b) SOP comparison on the Poincaré sphere of a sampled line of the vector fields of V-AO on and off status. Intensity comparison of a sampled line of different focus is presented as well. (c) V-AO off vs. V-AO on status of aberration correction for the GRIN objective lens. SOP distributions, phase patterns, and focus spot profiles are given.

The MM polarimeter provided complete information about the spatially variant polarising properties of the aberrating object and this information was used to apply the SOP correction in method A. However, it is also possible – and indeed sometimes more practical for various applications [2, 3] – to infer the necessary vectorial correction using less direct methods that do not require full knowledge of the polarisation aberration. We investigate here these polarimeter-free methods through the concept of "sensorless V-AO", which is named in analogy to wavefront-sensorless AO, which performs phase correction without the use of a wavefront sensor [2, 3]. In one case, a sequence of intensity measurements is taken in a conjugated pupil plane using a simple system consisting of polariser and a camera ("quasi-sensorless V-AO"; Method B); in the other case, a sequence of intensity measurements is taken at the focal plane instead of the pupil ("modal-sensorless V-AO"; Method C); see **Supplementary Note 4**. In both cases, a process is employed in order to optimize the intensity as a proxy for the uniformity of the polarisation state.

**Quasi-sensorless vectorial adaptive optics (Method B)**

Figure 3a gives a schematic of method B. At the heart of this method is the maximisation of intensity measured after a polariser across every point in the beam profile (highest intensity (HI); Fig. 3a). The camera is located at the pupil plane and measures the intensity of the output field after projection through a polariser, which was chosen to have the same principal polarisation eigenvector as the desired state (in principle, this could be spatially variant, but here linear polarisation is chosen for simplicity). When the output SOP is parallel to the eigenvector of the polariser, the intensity is maximum. One can thus correct the SOP aberration by maximizing the intensity at each point in the pupil.

For each point in the pupil, there is hence an input Stokes vector SV' that maximises the detected intensity and any other state will result in a reduced intensity. This intensity can be mapped onto the surface of the Poincaré sphere (see Fig 3a). The point with maximum intensity corresponds to the eigenvector of the analyser and hence indicates the optimal SV' pre-correction that must be applied with the V-AO module. In practice, this optimal correction can be obtained by sampling a number of trial configurations of the retardance settings of the two SLMs and by interpolating between these measurements to reach the maximum intensity (see exemplar in Fig 3a). This process can also be carried out in parallel for each single point in the pupil. Detailed algorithms are elaborated in **Supplementary Note 5**. This procedure to optimise the final SV correction introduces additional phase errors due to the SLMs. This necessitates the second correction step consisting of a conventional phase sensorless AO procedure.

We validated the experimental feasibility of method B using a titled waveplate array as the aberrating object. Such a waveplate array is widely used to generate vector vortex beams, such as for the depletion focus in a STED microscope [23]. We show the state before (V-AO off) and after correction (V-AO on) including the SOP fields, the phase on the DM, the focal spots (Fig. 3b). The improvement in performance is also exemplified through comparison of focal spot intensity profiles. A further demonstration showed compensation of the deleterious vectorial effects of a series of protected silver mirrors. These standard components of various modern optical systems show problematic vectorial aberrations, due to off-axis reflections, Fresnel's effects, and coating properties [6]. We applied V-AO correction through method B, providing the results of Fig. 3c. More detailed analysis of these and other vectorial aberrations due to other optical components, such as beam splitters, wavelength filters, etc., are given in **Supplementary Note 6**. We emphasise that the aberrations corrected here are spatially variant and thus require the complex correction provided by V-AO.

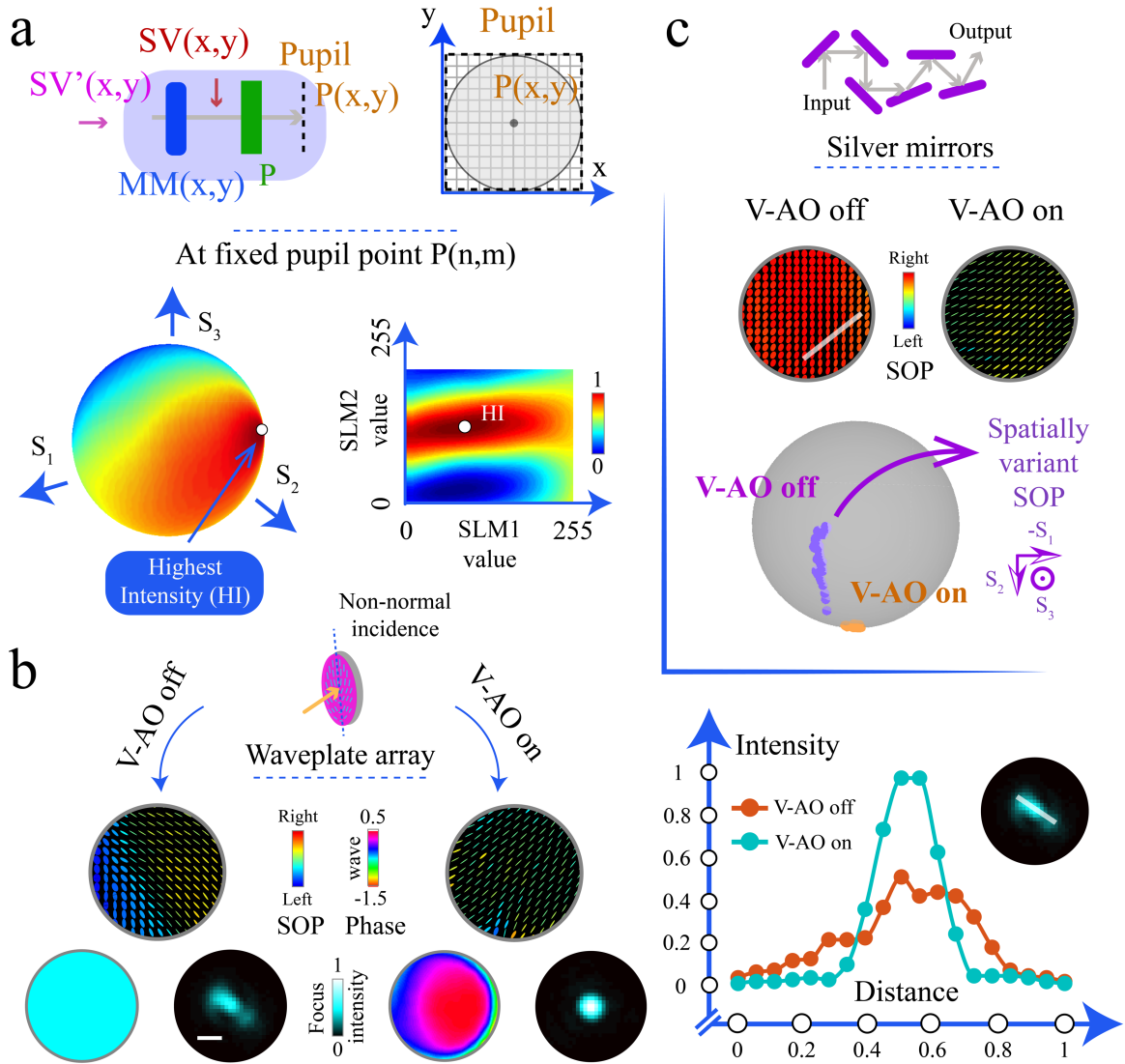

**Figure 3: Method B – quasi-sensorless vectorial adaptive optics.** (a) Optimisation of SV'(x,y), which represents a Stokes vector that varies across the pupil, which is located before the aberration described by the Mueller matrix MM(x, y) and is thus related to the transformed Stokes vector SV(x, y). For a fixed point in the pupil P(n, m), the intensity is mapped onto the Poincaré sphere as function of SV'(n, m). The intensity at the same point is equivalently mapped onto a graph of the SLM pixel values. The correction mechanism involves interpolation to the maximum intensity point through a sequence of measurements using different SV'(m,n). The detailed mechanism is elaborated in Supplementary Note 5. (b) V-AO off vs. V-AO on status with a tilted waveplate array. SOP distributions, DM phase patterns, foci profiles, and an exemplar for comparison are given. (c) Comparison of V-AO off and V-AO on status for an assembly of protected silver mirrors. SOP distributions and Poincaré sphere comparisons are given, showing compensation of the significant errors introduced by the mirrors.

**Modal-sensorless vectorial adaptive optics (Method C)**

Method B was able to correct vectorial aberrations after a short sequence of measurements at the conjugated pupil (correction of SOP) and focus (correction of phase). In this section, we demonstrate fully "sensorless" V-AO through method C. This method achieves V-AO correction with no additional hardware through focal plane measurement alone, taking advantage of prior knowledge of the nature of the vectorial aberration. There are various scenarios where assumptions can reasonably be made about the eigenmode axes of a polarising object, For example, this may occur in stressed optics, such as an endoscopic lens, which exhibits azimuthally or radially distributed birefringence axes, as determined by the intrinsic stress direction [4]; alternatively, in biological samples such as muscle tissue, where the axes follows the stress direction or the alignment direction of the intrinsic fibres [24]. While such prior knowledge about symmetry is useful, the value of retardance is often still unknown, such as due to variation from different manufacturing processes for the lens, or the state of the biological tissue [4, 7, 24], so adaptive vectorial correction is needed. Method C operates in analogy to conventional, phase-only sensorless methods, whereby different trial aberration corrections are applied in sequence and the optimal value of the correction is deduced by assessing the image quality using an optimisation metric [2, 3]. The salient difference for method C is that polarisation and phase aberration compensation are simultaneously performed with both SLMs and DM operating in concert at each step. This is a vectorial extension of phase correction through sensorless adaptive optics.

Sensorless feedback correction of both polarisation and phase is a challenging task, because of the complex interplay between them [6] when compensation is implemented using a combination of SLMs (see **Supplementary Note 1**). Furthermore, as feedback was provided solely through intensity measurements, a sequence of measurements was required to infer the necessary information for correction. We hence considered two different scenarios for method C. In scenario 1, we dealt only with an unknown polarisation aberration, assuming that the traditional phase aberration was negligible (Fig. 4a and 4b). At each point of the pupil, we had prior knowledge of the polarisation eigenmode axis, which we denoted by the line Q/-Q through the Poincaré sphere. The input SV (and equivalently the correction targeted SOP) was represented by point A. As the retardance value was unknown, the corrected SV T, which should be generated via the V-AO module, must lie on the circle K, which passed through A and was centred on Q/-Q.

In practice, this sensorless optimisation was implemented using vectorial modes defined over the whole pupil, which were based upon Zernike polynomials [2, 3]. The correction mechanism relied upon maximisation of the focal intensity (the process is elaborated in **Supplementary Note 7**), in a process that was a higher-dimensional analogue to sensorless phase AO. Optimising the intensity through adjustment of the modes on the SLMs provided an indirect route to optimisation of the SOP. In scenario 2 (Fig. 4c), we considered the combination of unknown polarisation and phase aberrations. Here, we first executed a phase-only sensorless correction routine to remove most of the phase aberration due to optical path differences on propagation through the aberrating sample. Then we applied a polarisation-only aberration correction, following the procedure used in scenario 1. Finally, we applied a further round of phase-only aberration correction to remove any residual phase bias. In effect, through maximisation of focal intensity, this V-AO procedure seeks uniformity of the SOP, rather than a specific SOP (see details in **Supplementary Note 7**).

Experimentally, we validated the feasibility of our method C (sensorless scenario 1) via a calibration target consisting of a region of birefringent material made from a piece of thin film waveplate). This creates a pure polarisation aberration with negligible phase offset (Fig. 4b). Specifically, we present the final overall system axis orientation from the combined effect of the V-AO module and the object. This in effect determines the overall system performance, as an alternative evaluation of our V-AO performance. It can be seen that the effective fast/slow axis distribution show significantly improved uniformity after correction. More data, analysis and discussion are provided in **Supplementary Note 7**. V-AO correction for scenario 2 was undertaken using a thin tendon tissue sample, which exhibited both polarisation and phase effects, in this case with comparatively smaller polarisation aberrations (Fig. 4c). Focal spot performance and axis corrections are provided. Detailed analysis of the performance, including in terms of the vectorial modes, is provided in **Supplementary Note 7**. The above results validate that the sensorless V-AO method C has potential to be used in future applications that require precise vectorial control and correction of vectorial aberrations, where indirect optimisation is required, such as for beam shaping, focal control and imaging.

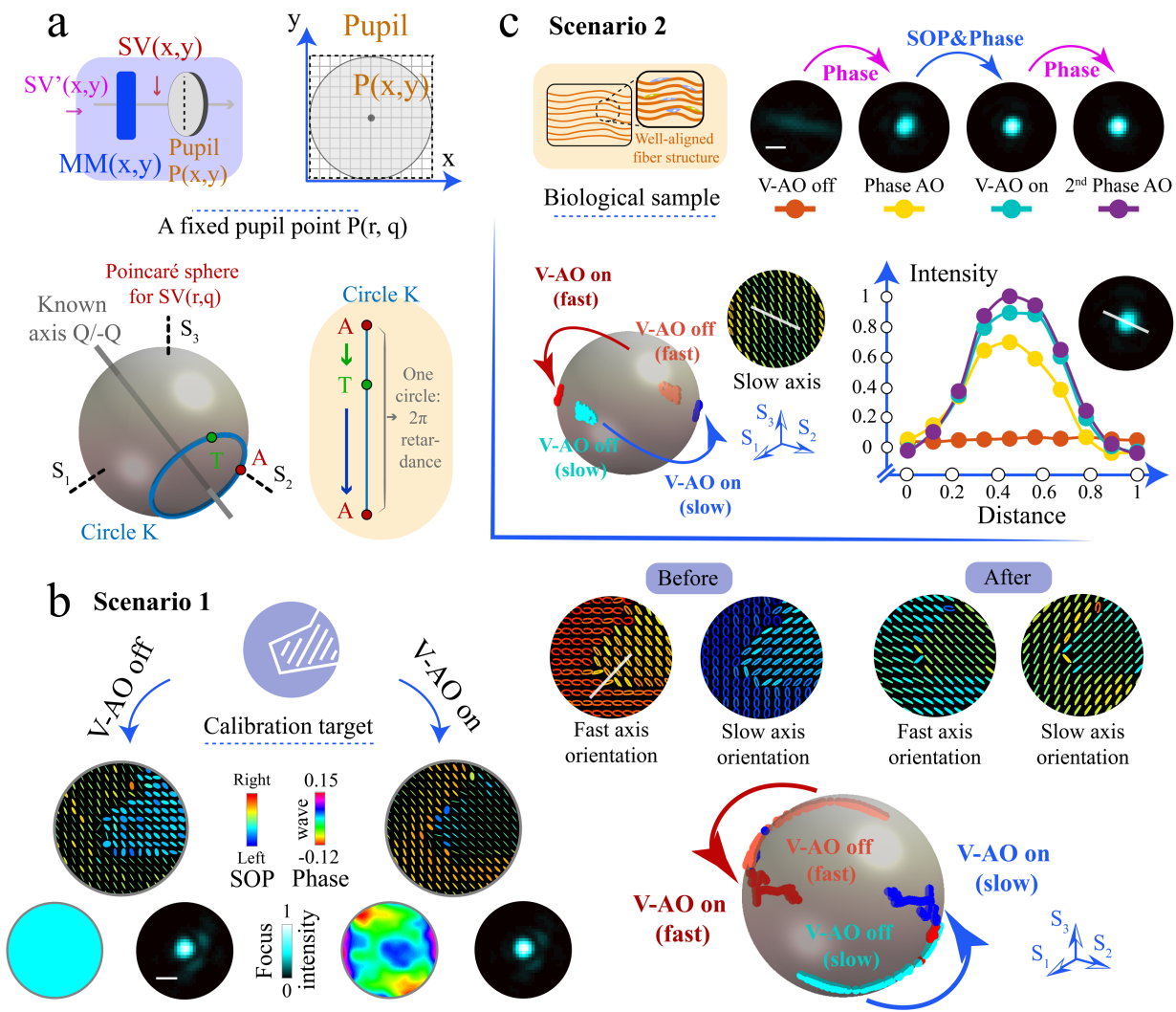

**Figure 4: Method C – sensorless vectorial adaptive optics.** (a) Simplified schematic of the sensorless V-AO mechanism. Taking a fixed point P(r,q) on the pupil as example, when the axis of the aberration corresponding to this point is known, the circle K is determined. Hence, the vectorial correction must be found to set the input SV'(r,q), corresponding to point T, back to the desired SV(r,q) at point A. In practice, aberration modes are applied that change all of the points on the pupil P(x,y) simultaneously to optimise the focal intensity. Details are elaborated in Supplementary Note 7. (b) V-AO off vs. V-AO on status for a calibration target consisting of piece of thin film waveplate. SOP distributions, patterns on DM, focus profiles, overall slow/fast axis performances (before and after the correction), as well as the axis distributions and their comparison on Poincaré sphere are given. The corrected SOP is more uniform, and the focal spot is sharper. The complex phase pattern here reveals the extra phase errors that introduced via AO correctors. It can be also observed that both corrected axes are more uniform. (c) Comparison of V-AO off and V-AO on status of a biological sample. Focal spots distributions at three steps, axis orientation corrections, are given for comparison. It can be found that the axis distribution and focal spot profiles are enhanced after the correction process.

**Discussion**

In summary, we have extended the concepts of conventional phase-only AO into the vectorial domain by merging polarisation and phase aberration correction. V-AO can be implemented using feedback in a sensor-based approach using a polarimeter, or through two different implementations of "sensorless V-AO", which uses neither a MM polarimeter nor a wavefront sensor.

These three approaches provide a versatile toolkit for compensation of vectorial aberrations, which can be chosen to match conditions in real applications. The sensor-based method provides the most comprehensive approach compatible with full MM characterisation, should it be necessary. However, the quasi-sensorless (and modal-sensorless) method can provide the ability to perform V-AO with simpler hardware requirements compatible with more practical scenarios [2, 3]. The quasi-sensorless method features unique advantages: First, the complex hardware necessary for a full MM polarimeter measurement is replaced by a simple analyser without moving parts. Second, calculation of the MM is no longer needed, nor is its decomposition for application of the SLM settings, which avoids the complex error amplification due to the MM matrix calculation [7]. Third, detailed calibration of the SLMs is not necessary, as the optimization procedure can be expressed directly in terms of the pixel values applied to the two SLMs. The modal-sensorless approach requires no additional hardware and no extra pupil measurement, uses prior knowledge of the target, and is based purely on focal optimisation. These methods can separately assist different optical system scenarios, with intriguing directions that are largely unexplored.

Future developments can expand these methods to deal with light depolarised by the object, which can occur in scenarios such as bulk tissue monitoring or diagnosis [7]. Furthermore, with an increased number of AO devices, the V-AO format can be extended. The current system, using two SLM passes, can provide vectorial aberration compensation with pre-correction of the illumination state. For even wider reach, the system could be expanded to three SLM passes [17], thus permitting conjugation of the vectorial state, for example in the emission path of a microscope. More discussion and perspectives can be found in **Supplementary Notes 8**.

Overall, we have put forward novel V-AO techniques for joint correction of polarisation and phase aberrations. With these vectorial field feedback control methods, this next-generation AO technique is expected to benefit various research areas – for example [25-30], ranging from astronomical telescopes to microscopy – with further applications spanning across galaxy detection, to laser-based and lithographic nano-fabrication, in addition to biomedical and clinical characterization.

**Acknowledgments**

The project was supported by the European Research Council (AdOMiS, no. 695140).

**Competing interests**

The authors declare no competing interests.

**Author contributions**

C.H., J.A. and M.J.B. originated the project and conceived the V-AO approaches. C.H. designed and built the experimental hardware, conducted polarimetry software, prepared the samples. C.H. and J.A. planned the simulation, implementation of both the sensor and sensorless algorithm, and conducted the experiments together. J.A. performed simulation and algorithm, control software and data extraction methods. J.A., C.H. and M.J.B. analysed the data, wrote and edited the paper. C.H. prepared the figures. M.J.B. oversaw the project.

**Additional Information**

Correspondence and request for materials should be addressed to C.H., J.A., or M.J.B.

# Vectorial adaptive optics
# – Supplementary information –

Chao He[†,*], Jacopo Antonello[†,*], and Martin J. Booth[*]

[*]Department of Engineering Science, University of Oxford, Parks Road, Oxford, OX1 3PJ, UK

[†]These authors contributed equally to this work
[*]Corresponding authors: chao.he@eng.ox.ac.uk; jacopo.antonello@eng.ox.ac.uk; martin.booth@eng.ox.ac.uk

## Supplementary Note 1:

## Mathematical background

We can model the evolution of the state of polarisation (SOP) and the scalar phase throughout the optical system using conventional Jones calculus [1–3]. For a single point in the pupil, the change in SOP and phase can be modelled as

$$\boldsymbol{j}_2 = J\boldsymbol{j}_1, \tag{1}$$

where $\boldsymbol{j}_1$ and $\boldsymbol{j}_2$ are the Jones vectors respectively before and after the change in SOP and scalar phase applied by a polarisation modulation element such as a wave plate. This latter is modelled using a a $2 \times 2$ Jones matrix $J$. It is useful to parametrise $J$ so that the effects of SOP and phase change are more easily identifiable. In this paper we only consider retardance and phase modulation for a fully-polarised beam of light. As a result, $J$ can be expressed as

$$J = e^{i\phi}U, \tag{2}$$

where $\phi$ is the scalar phase applied by the element and $U$ is a special unitary matrix (SU) describing the change in SOP. This latter matrix can also be parametrised [4] as

$$U = \mathrm{SU2}(\boldsymbol{Q}, \theta) = \cos(\theta/2)I + i\sin(\theta/2)(n_1\sigma_1 + n_2\sigma_2 + n_3\sigma_3), \tag{3}$$

where $\theta$ is the retardance applied by the element, $I$ is the $2 \times 2$ identity matrix, and $\sigma_1$, $\sigma_2$, $\sigma_3$ are the Pauli matrices defined as

$$\sigma_1 = \begin{bmatrix} 1 & 0 \\ 0 & -1 \end{bmatrix}, \sigma_2 = \begin{bmatrix} 0 & 1 \\ 1 & 0 \end{bmatrix}, \sigma_3 = \begin{bmatrix} 0 & i \\ -i & 0 \end{bmatrix}. \tag{4}$$

Note that $\sigma_3$ is defined differently than in some other sources [1–3,5,6]. The coefficients $n_1$, $n_2$, $n_3$ belong to a vector $\boldsymbol{Q} = [n_1; n_2; n_3]$ with unit norm, i.e., $|\boldsymbol{Q}| = 1$. The change in SOP modelled by $U$ can also be visualised as a rotation in the Poincaré sphere, where $\boldsymbol{Q}$ is the axis of rotation and $\theta$ the angle of rotation. Given a Jones vector $\boldsymbol{j}$, the normalised Stokes vector representing its SOP can be computed as [6]

$$\boldsymbol{S} = \begin{bmatrix} \boldsymbol{j}^\dagger \sigma_1 \boldsymbol{j} \\ \boldsymbol{j}^\dagger \sigma_2 \boldsymbol{j} \\ \boldsymbol{j}^\dagger \sigma_3 \boldsymbol{j} \end{bmatrix}, \tag{5}$$

where $\cdot^\dagger$ denotes the conjugate transpose. Note that by choosing $\sigma_3$ as in Eq. (4), we have that left circular (LC) polarisation $\boldsymbol{j}_{LC} = [1;i]$ corresponds to the South pole $\boldsymbol{S}_{LC} = [0;0;-1]$ in the Poincaré sphere. Right circular (RC) $\boldsymbol{j}_{RC} = [1;-i]$ corresponds to the North pole $\boldsymbol{S}_{RC} = [0;0;1]$, instead. This is in accordance with the convention most commonly employed in optics [7] where RC polarised light is represented by the Jones vector $[1;-i]$ and corresponds to the North pole.

Positive uniaxial elements such as liquid crystals (LC) in a spatial light modulator (SLM) are characterised by an ordinary $n_o$ and extraordinary $n_e$ indices of refraction. The magnitude of the effective $n_e$ of a LC cell can be controlled by applying a voltage. Excluding a common phase delay applied to both the $n_o$ and $n_e$ axes, the effect of a LC cell can be modelled by

$$J_{LC} = \begin{bmatrix} e^{i\phi} & 0 \\ 0 & 1 \end{bmatrix}, \tag{6}$$

where $\phi$ represents the additional phase delay experienced along the extraordinary axis. Note that in Eq. (6) we have dropped a common phase factor that would account for the propagation along a non-birefringent cell with uniform refractive index $n_o$. This factor is inconsequential for our analysis as we are only concerned with the additional optical path length (OPL) experienced along the $n_e$ axis, which is accounted for by $\phi$, with respect to the OPL experienced along the $n_o$ axis. It can be seen that

$$J_{LC} = e^{i\theta/2}\begin{bmatrix} e^{i\theta/2} & 0 \\ 0 & e^{-i\theta/2} \end{bmatrix} = e^{i\theta/2} \cdot \text{SU2}(\boldsymbol{H},\theta), \tag{7}$$

where $\boldsymbol{H} = [1;0;0]$ denotes the horizontal axis in the Poincaré sphere. Therefore, Eq. (7) can be interpreted as applying a scalar phase factor of $\theta/2$ to both the first and second components of the Jones vector. In addition to that, a change in the SOP is also applied via $\text{SU2}(\boldsymbol{H},\theta)$. This latter can be visualised as a rotation of an angle $\theta$ about the $\boldsymbol{H}$ axis in the Poincaré sphere. We remark that, contrary to the case of Eq. (6), in Eq. (7) we choose to keep the phase factor $e^{i\theta/2}$. This is because Eq. (7) only applies to a single point of the cell, but $\theta$ can have different values for each pixel of a LC cell. As a result, a phase profile develops across the cell if one assigns different values to $\theta$ for different pixels of the LC cell.

## 1.1 Modelling the SOP and phase in the optical system

The layout of the optical system is shown in Supplementary Figure 1, where different positions along the setup are enumerated by numbers in purple. We first consider a single point in the pupil of the optical system. Given an initial Jones vector $\boldsymbol{j}_1$ at position (1), the Jones vector after reflection off the first SLM at position (2) is found to be

$$\boldsymbol{j}_2 = e^{i\theta_1/2} J_r \, \text{SU2}(\boldsymbol{H},\theta_1)\boldsymbol{j}_1. \tag{8}$$

Here, $J_r = \sigma_1$ accounts for the reflection at the back plane of the SLM, see Chapter 17 in [1]. Note that $J_r = -i\,\text{SU2}(\boldsymbol{H},\pi)$, and therefore can also be thought of as a rotation by 180° about the $\boldsymbol{H}$ axis in the Poincaré sphere. The retardance applied with the first SLM is $\theta_1$. We remark that in reality the retardance $\theta_1$ is applied in two steps. First, a retardance of $\theta_1/2$ is applied when the beam passes through the LC cell in the forward direction. This is followed by the reflection off the back plane and by another pass through the LC cell in the backward direction, which applies another retardance of $\theta_1/2$. Since in our modelling the axes of the LCs are parallel to the $x$ axes in both the forward and backward frames of reference, this description is equivalent to Eq. (8).

After reflection off the first SLM the beam is incident onto a half-wave plate oriented as shown in Supplementary Figure 1(a). As a result, the Jones vector after transmission through the half-wave plate at position (3) is given by

$$\boldsymbol{j}_3 = e^{i\pi/2}\,\text{SU2}(\boldsymbol{W},\pi)\boldsymbol{j}_2, \tag{9}$$

where $\boldsymbol{W} = [1; -1; 0]/\sqrt{2}$ is the axis of rotation in the Poincaré sphere. The reflection off the second SLM at position (4) can be modelled as

$$\boldsymbol{j}_4 = e^{i\theta_2/2} J_r \, \text{SU2}(\boldsymbol{H}, \theta_2) \boldsymbol{j}_3, \qquad (10)$$

where $\theta_2$ is the retardance applied with the second SLM. At position (5), scalar phase modulation is applied with a deformable mirror (DM), which is modelled by the Jones matrix $e^{i\psi} I$, i.e.,

$$\boldsymbol{j}_5 = e^{i\psi} I \boldsymbol{j}_4. \qquad (11)$$

At position (6), the beam is perturbed by an aberrating object, which comprises a scalar phase $\phi$ and an aberration of the SOP, which is parametrised by an axis $\boldsymbol{Q}$ and a retardance $\delta$, i.e.,

$$\boldsymbol{j}_6 = e^{i\phi} \, \text{SU2}(\boldsymbol{Q}, \delta) \boldsymbol{j}_5. \qquad (12)$$

Depending on the considered scenario, $\phi$, $\boldsymbol{Q}$, and $\delta$ may be unknown and to be determined by one of the correction methods outlined in the sections below. The scalar phase $\phi$ accounts for OPL differences accrued by propagation of the beam through the aberrating object. These are commonly compensated with conventional adaptive optics (AO) [8]. Finally, overall, the input and output Jones vectors $\boldsymbol{j}_1$ and $\boldsymbol{j}_6$ are thus related by

$$\begin{aligned}
\boldsymbol{j}_6 = {}& \exp\{i[\psi + \phi + (\theta_1 + \theta_2)/2]\} \\
& \times \text{SU2}(\boldsymbol{Q}, \delta) \\
& \times \text{SU2}(\boldsymbol{H}, \pi) \, \text{SU2}(\boldsymbol{H}, \theta_2) \\
& \times \text{SU2}(\boldsymbol{W}, \pi) \\
& \times \text{SU2}(\boldsymbol{H}, \pi) \, \text{SU2}(\boldsymbol{H}, \theta_1) \\
& \times \boldsymbol{j}_1.
\end{aligned} \qquad (13)$$

Note that in the equation above we have dropped some constant phase factors that do not vary across the pupil and are thus inconsequential for our analysis. We emphasise that Eq. (13) is only relevant for a single point in the pupil disk. In fact, when spatial variations are considered, all the Jones vectors, $\psi$, $\phi$, $\theta_1$, and $\theta_2$ should be replaced by functions of the coordinates in the pupil plane. Finally, Eq. (13) also highlights the total accumulated scalar phase i.e, $\psi + \phi + (\theta_1 + \theta_2)/2$, as well as a number of rotations on the Poincaré sphere.

We can use Eq. (13) above to establish under which conditions both the SOP and the phase can be corrected in the pupil of the overall optical system. For this reason, we first assume that $\boldsymbol{Q}$, $\delta$, and $\phi$ are known for each point in the pupil. The SOP aberration of the object is determined by $\boldsymbol{Q}$ and $\delta$, whereas the scalar phase aberration is determined by $\phi$. Note that, in general, it is not possible to select $\theta_1$ and $\theta_2$ such that the matrix products in Eq. (13) evaluate to the identity matrix $I$. That would imply that two SLMs can make an arbitrary retarder and correct for all possible values of $\boldsymbol{j}_1$. Instead, the two SLMs can only make a correction for a single, fixed value of $\boldsymbol{j}_1$, which we have chosen to be $\boldsymbol{j}_1 = [1; -1]/\sqrt{2}$. This is because two SLMs are insufficient to simulate an equivalent arbitrary wave plate, which can be realised by employing three SLMs in series – see [9]. Nevertheless, correction of the SOP can still be achieved by choosing $\theta_1$ and $\theta_2$ such that $\boldsymbol{S}_6 = \boldsymbol{S}_1$, where $\boldsymbol{S}_1$ and $\boldsymbol{S}_6$ are the Stokes vectors on the Poincaré sphere corresponding to $\boldsymbol{j}_1$ and $\boldsymbol{j}_6$, respectively. This condition is satisfied when

$$\text{SU2}(\boldsymbol{Q}, \delta) \, \text{SU2}(\boldsymbol{H}, \pi) \, \text{SU2}(\boldsymbol{H}, \theta_2) \, \text{SU2}(\boldsymbol{W}, \pi) \, \text{SU2}(\boldsymbol{H}, \pi) \, \text{SU2}(\boldsymbol{H}, \theta_1) = \text{SU2}(\boldsymbol{S}_1, \xi), \qquad (14)$$

i.e., when the product of the matrices in Eq. (13) evaluates to a unitary matrix $\text{SU2}(\boldsymbol{S}_1, \xi)$ for which $\boldsymbol{j}_1$ is an eigenvector. In this case, $\xi$ represents the geometric phase accumulated by the closed path travelled along the Poincaré sphere [10, 11]. Once Eq. (14) is satisfied, correction of the total scalar phase is achieved by letting

$$\psi = -[\phi + (\theta_1 + \theta_2)/2 + \xi]. \qquad (15)$$

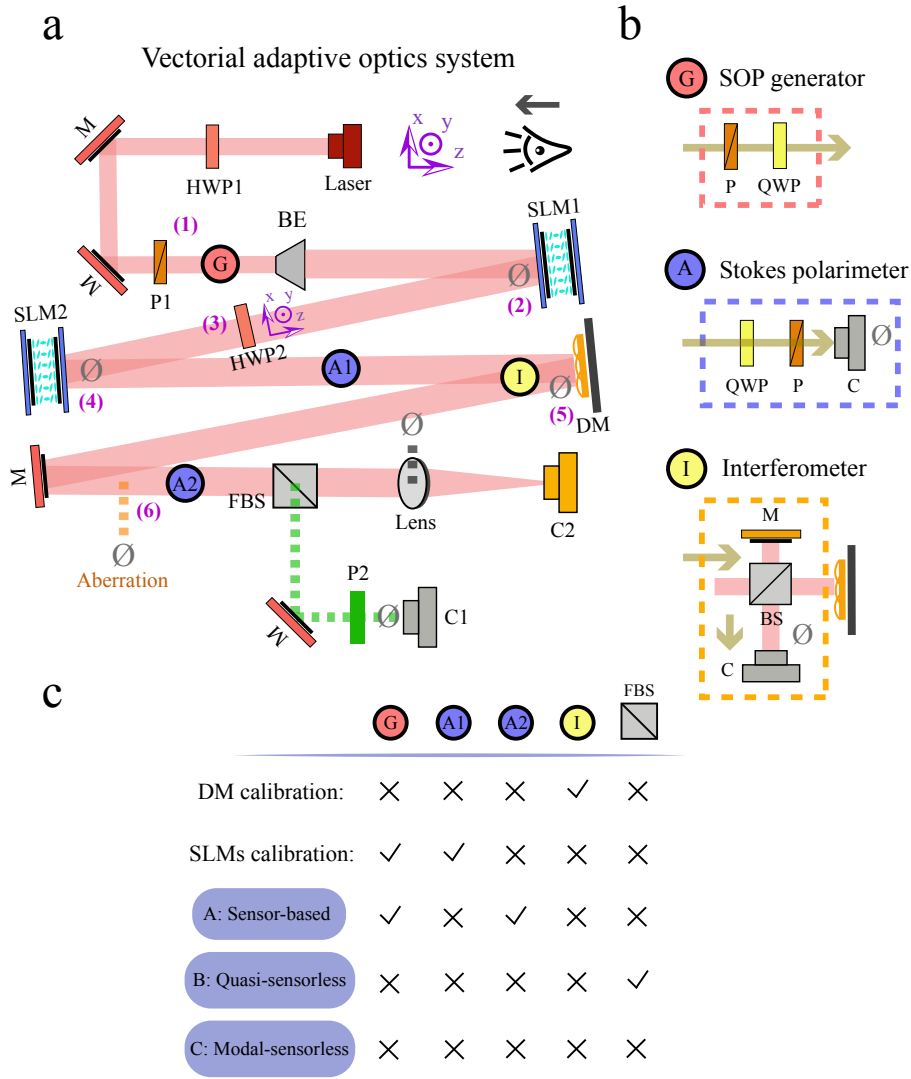

**Supplementary Figure 1:** Simplified layout of the optical system used in the experimental validation. (a) A HeNe laser beam (Melles Griot, 05-LHP-171, 632.8 nm) is expanded and directed towards the first SLM (SLM1; Hamamatsu, X10468-01). After reflection the beam passes through a half-wave plate (HWP2; Thorlabs, WPH10M-633) and is reflected off the second SLM (SLM2; Hamamatsu, X10468-01) first, and then off a DM (Boston Micromachines Corporation, Multi-3.5). The planes conjugate to the pupil plane are denoted with ∅ (the telescopes that reimage these planes are not depicted for simplicity). Some positions in the setup are enumerated with numbers between parentheses in purple. The frame of reference at position (3) is used to indicate the orientation of HWP2. Measuring positive angles from the $x$ towards the $y$ axis, the slow and fast axes of HWP2 are at 22.5° and 112.5°, respectively. Legend: HWP half-wave plate, M flat mirror, P polariser (Thorlabs, GL10-A), BE beam expander, BS beam splitter (Thorlabs, BS010), FBS flip-in beam splitter (Thorlabs, BS010), C camera (Thorlabs, DCC3240N), QWP quarter-wave plate (Thorlabs, WPQ10M-633); (b) Flip-in elements denoted by letters in (a); (c) Table showing in which cases the flip-in elements depicted (b) are enabled.

The modelling discussed so far was validated with experimental measurements in Supplementary Figure 2, where the first row reports the results of numerical simulation, whereas the second row reports experimental measurements obtained with the setup depicted in Supplementary Figure 1. In this case, no aberrating object was inserted at position (6) in Supplementary Figure 1. Instead, a randomly chosen aberration was induced with the two SLMs only. In Supplementary Figure 2 a comparison is made between the Mueller matrices (MM) computed using the model above shown in (a) and the effective MM measured experimentally shown in (d). Supplementary Figure 2(b) shows the computed SOP expected at position (6). The same data is obtained using the experimentally measured MM and is shown in Supplementary Figure 2(e). Finally, we computed the expected focal intensity distribution (FID), which is reported in Supplementary Figure 2(c). The corresponding experimental measurement obtained with camera C2 is reported in Supplementary Figure 2(f). The experimental MMs were obtained by applying the polarimetry method described in [12] and by temporarily enabling the SOP generator G and Stokes polarimeter A1 depicted in Supplementary Figure 1.

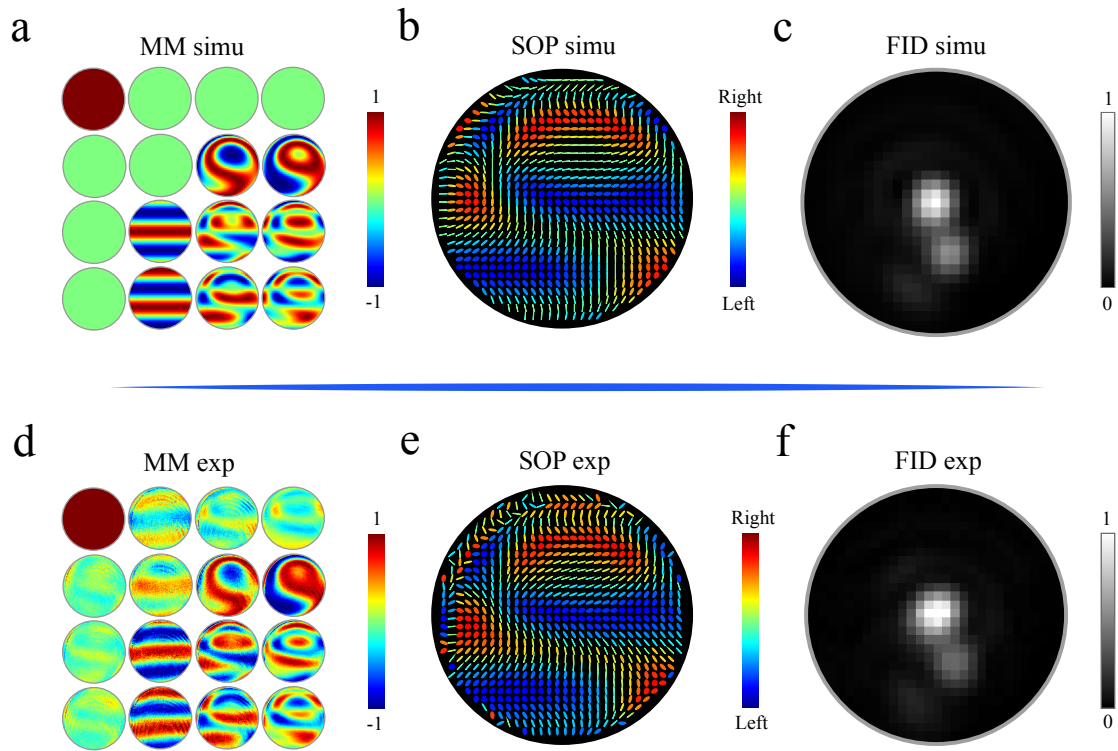

**Supplementary Figure 2:** Experimental validation of the modelling reported in Supplementary Note 1.1. A randomly chosen aberration is induced with both SLMs and no aberrating object is present at position (6) in Supplementary Figure 1. (a) MM data computed using the modelling relating position (1) and position (6); (b) SOP predicted at position (6) using the modelling when $\bm{j}_1 = [1; -1]/\sqrt{2}$; (c) FID computed using the modelling and expected at camera C2; (d) Experimental MM data measured with MM polarimetry [12]; (e) SOP predicted at position (6) using the experimentally measured MM data; (f) FID measured with camera C2.

## 1.2 Arbitrary modulation of the SOP

As seen above, the Jones vector $j_4$ after reflection off the second SLM is given by

$$\begin{aligned} j_4 = {} & \exp\{i(\theta_1 + \theta_2)/2]\} \\ & \times \mathrm{SU2}(\boldsymbol{H}, \pi)\,\mathrm{SU2}(\boldsymbol{H}, \theta_2) \\ & \times \mathrm{SU2}(\boldsymbol{W}, \pi) \\ & \times \mathrm{SU2}(\boldsymbol{H}, \pi)\,\mathrm{SU2}(\boldsymbol{H}, \theta_1) \\ & \times j_1. \end{aligned} \quad (16)$$

From now onwards, we always assume that $j_1$ is fixed to $j_1 = [1;-1]/\sqrt{2}$, so that the corresponding SOP at position (1) is fixed to $\boldsymbol{S}_1 = [0;-1;0]$ – as can be found by applying Eq. (5). Next, we consider the problem of selecting $\theta_1$ and $\theta_2$ such that the SOP $\boldsymbol{S}_4$ at position (4) can be set to a desired value, i.e., $\boldsymbol{S}_4 = [x_4; y_4; z_4]$. By reasoning out the rotations along the Poincaré sphere outlined in Eq. (16), it can be shown that this is achieved when

$$\begin{cases} \theta_1 &= \arccos(-x_4) \\ \theta_2 &= \mathrm{arctan2}(-y_4, z_4). \end{cases} \quad (17)$$

As a result, the two SLMs are able to transform the uniform SOP given by $\boldsymbol{S}_1 = [0;-1;0]$ into an arbitrary, spatially variant SOP at position (4). This means that the SOP can be pre-corrected before propagating through the aberrating object at position (6), see Supplementary Figure 1.

An illustration outlining the rotations in the Poincaré sphere is provided in Supplementary Figure 3. Here we have chosen $\boldsymbol{S}_4 = [0.6943; -0.5826; 0.4226]$, from which we obtain $\theta_1 = 134.0°$ and $\theta_2 = 54.0°$ after applying Eq. (17). First, retardance $\theta_1$ is applied with SLM1, which causes a rotation of 140.0° about the $\boldsymbol{H}$ axis. This is followed by a reflection off the back plane of the SLM, which is modelled by a rotation of 180.0° about the same axis $\boldsymbol{H}$. As a result, the SOP $\boldsymbol{S}_2$ is reached, see Supplementary Figure 3(b). At this point, the beam passes through the half-wave plate, which is modelled by a rotation of 180.0° about the $\boldsymbol{W}$ axis. This results in a SOP of $\boldsymbol{S}_3$, see Supplementary Figure 3(c). Finally, retardance $\theta_2$ is applied, which causes a rotation of 54.0° about the $\boldsymbol{H}$ axis. Another rotation about the $\boldsymbol{H}$ axis models the reflection off the back plane of the SLM. This results in the desired state $\boldsymbol{S}_4$ being reached as expected, see Supplementary Figure 3(d).

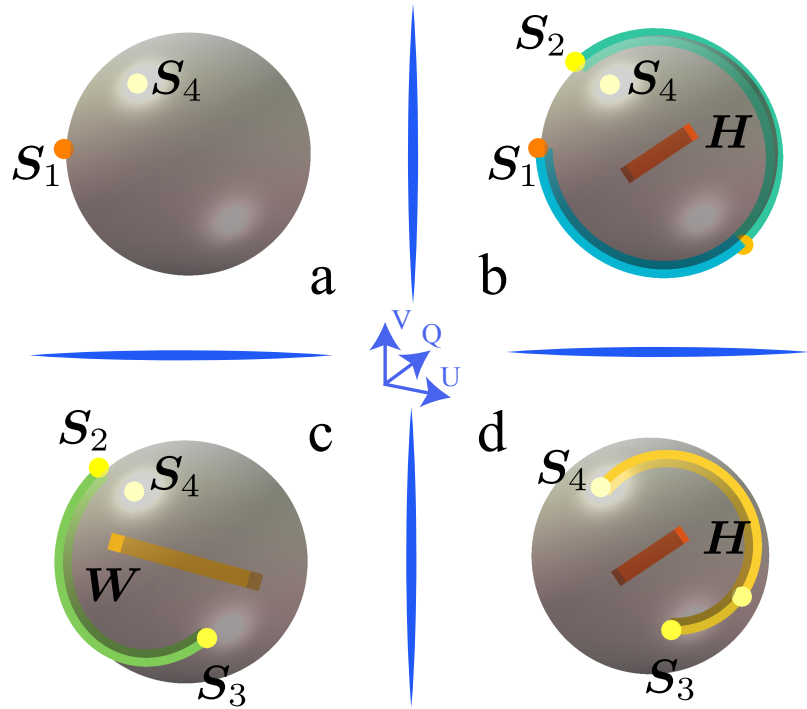

**Supplementary Figure 3:** Generation of an arbitrary SOP at position (4) in the optical setup. The Stokes vectors $S$ indicate the SOP at various positions in the setup, which are denoted by the number in the subscript. (a) The initial state $S_1$ is fixed to $[0; -1; 0]$, whereas an arbitrary state $S_4$ is desired; (b) Retardance $\theta_1$ is applied with the first SLM, which is modelled as a rotation by $\theta_1$ along the $H$ axis. This is followed by a rotation of 180° modelling the reflection off the back plane of the SLM; (c) The effect of the half-wave plate is that of applying a rotation of 180° about the $W$ axis; (d) The desired state $S_4$ is reached using two rotations about the $H$ axis modelling both the retardance $\theta_2$ applied with SLM2 and the reflection off the back plane.

## 1.3 Conversion between Mueller and Jones matrices

Given a $2 \times 2$ complex-valued Jones matrix $J$, the corresponding $4 \times 4$ real-valued Mueller matrix $M$ can be found as a function of the Pauli matrices as shown below. Let

$$P = \begin{bmatrix} \text{vec}(I) & \text{vec}(\sigma_1) & \text{vec}(\sigma_2) & \text{vec}(\sigma_3) \end{bmatrix}, \tag{18}$$

and

$$M = \frac{1}{2} P^\dagger \begin{bmatrix} \text{vec}(J^\dagger J) & \text{vec}(J^\dagger \sigma_1 J) & \text{vec}(J^\dagger \sigma_2 J) & \text{vec}(J^\dagger \sigma_3 J) \end{bmatrix}, \tag{19}$$

where $\text{vec}(\cdot)$ denotes the vectorisation operation. Note that by choosing $\sigma_3$ as in Eq. (4), the convention for RC and LC employed in optics literature is used [7]. Other conventions, such as the one employed in [3], can be employed if $\sigma_3$ is replaced by its conjugate transpose.

We now consider the inverse problem of determining $J$ from a given $M$. This latter problem has multiple solutions, since different Jones matrices expressing different global phases can correspond to the same Mueller matrix. Numerous authors have provided formulas for the individual elements of $J$, see for example Appendix D in [3] and Chapter 6 in [1]. These formulas are nonetheless inconvenient to use as special care must be taken for the cases when some denominators vanish. In the following, instead, we propose to formulate the computation of $J$ as a Frobenius norm minimisation problem [13], i.e.,

$$\min_{J^T, J^\dagger \in \mathbb{C}^2} \left\| \frac{1}{2} P M^T P^\dagger - J^T \otimes J^\dagger \right\|_F = \\ \left\| K - J^T \otimes J^\dagger \right\|_F = \\ \left\| R - \text{vec}(J^T) \text{vec}(J^\dagger)^T \right\|_F, \tag{20}$$

where $\|\cdot\|_F$ is the Frobenius norm, $K = 1/2 \cdot P M^T P^\dagger$, $\otimes$ denotes the Kronecker product, and

$$R = \begin{bmatrix} K_{1,1} & K_{2,1} & K_{1,2} & K_{2,2} \\ K_{3,1} & K_{4,1} & K_{3,2} & K_{4,2} \\ K_{5,1} & K_{6,1} & K_{5,2} & K_{6,2} \\ K_{1,3} & K_{2,3} & K_{1,4} & K_{2,4} \\ K_{3,3} & K_{4,3} & K_{3,4} & K_{4,4} \\ K_{5,3} & K_{6,3} & K_{5,4} & K_{6,4} \end{bmatrix}. \tag{21}$$

The optimisation in Eq. (20) can be solved [13] by taking the SVD of $R = \bar{U} \Sigma \bar{V}^\dagger$ and letting $J^T = \text{vec}^{-1}(\sqrt{\Sigma_{1,1}} \bar{u}_1)$, where $\Sigma_{1,1}$ is the first singular value in $\Sigma$ and $\bar{u}_1$ is the first column of $\bar{U}$. In this way, all the elements of $J$ can be computed simultaneously using only matrix-based operations, in contrast to conventional formulas, e.g., see [1, 3].

# Supplementary Note 2:
# Calibration of the SLMs

In this section we describe a procedure to calibrate the SLMs by experimentally determining the relationship between the pixel value and the corresponding retardance effectively induced by each SLM. Initially, we consider SLM1 in Supplementary Figure 1. The calibration procedure involves collecting two sets of data using polariser P2 and camera C1 in Supplementary Figure 1. For each set of data, the pixel value is increased from the minimum up to its maximum and an image is recorded at each step. For the first dataset, we set the input SOP at position (1) to $S_1 = [0; -1; 0]$ and the SOP transmitted by polariser

P2 to $\boldsymbol{S}_a = [1;0;0]$. The SOPs are defined with respect to the frame of reference depicted in purple in the top-right hand corner of Supplementary Figure 1(a). Using Mueller calculus and the equations from Supplementary Note 1.1, it can be shown that the intensity recorded with camera C1 varies as

$$I_x = \bar{I}_x(1-\cos\theta_1)/2, \tag{22}$$

where $\bar{I}_x$ is a scaling parameter accounting for the maximum intensity. Note that Eq. (22) holds independently of what retardance SLM2 is applying. This results in the following set of input–output measurements

$$\mathcal{S}_1 = \{(g_{1,1}, I_{x,1}), \ldots, (g_{1,L}, I_{x,L})\}. \tag{23}$$

for each pixel within the aperture disk. In our case, we have that $g_1$ can vary between $g_{1,1} = 0$ and $g_{1,L} = 255$.

For the second dataset, instead, we temporarily set the input SOP at position (1) to RC, i.e., $\boldsymbol{S}_1 = [0;0;1]$, and keep the SOP of polariser P2 fixed to $\boldsymbol{S}_a = [1;0;0]$. Again, using Mueller calculus, one can show that the intensity recorded by camera C1 varies as

$$I_y = \bar{I}_y(1-\sin\theta_1)/2. \tag{24}$$

The second dataset is therefore given by

$$\mathcal{S}_2 = \{(g_1, I_{y,1}), \ldots, (g_L, I_{y,L})\}. \tag{25}$$

By looking at Eq. (22) and Eq. (24) it can be seen that one can estimate the retardance $\theta_1$ modulo $2\pi$ with

$$\hat{\theta}_1 = \arctan2(1 - 2I_y/\bar{I}_y, 1 - 2I_x/\bar{I}_x). \tag{26}$$

This latter equation can therefore be applied to the measurements in $\mathcal{S}_1$ and $\mathcal{S}_2$, obtaining the desired mapping between $g_1$ and the effective retardance $\theta_1$. Finally, one can use the obtained mapping to build a lookup table, so that the necessary value of $g_1$ can be determined for a desired retardance value of $\theta_1$. Note that by applying this recipe, no phase unwrapping [14] is required if the full range of $\theta_1$ is within a single period. This procedure allows us to obtain a different mapping for each pixel of SLM1 that falls within the aperture. In our case, we averaged the mappings obtained for each pixel to retrieve a single mapping approximately valid for the full aperture. Nevertheless, a more refined approach could be implemented whereby a pixel-by-pixel calibration is performed instead.

We emphasise that this calibration procedure is advantageous with respect to previously proposed methods that rely on a single dataset only. For example, in [15], the mapping between $g_1$ and $\theta_1$ is obtained by fitting Eq. (22) to the data of $\mathcal{S}_1$ only. Nevertheless, one has that the cosine function is only injective over half a period. As a result, it is not possible to tell from the data in $\mathcal{S}_1$ alone whether the argument of the cosine is either $\theta_1$ or $2\pi - \theta_1$, since $\cos\theta_1 = \cos(2\pi - \theta_1)$ for each element in $\mathcal{S}_1$. Even if phase unwrapping [14] is applied along the elements of $\mathcal{S}_1$, multiple solutions to the fitting of Eq. (22) exist. This ambiguity is resolved by considering the corresponding measurements in $\mathcal{S}_2$.

The same procedure can be applied to calibrate the second SLM with some minor modifications as follows. For the first dataset, we temporarily set the input SOP at position (1) to $\boldsymbol{S}_1 = [1;0;0]$ and the analyser SOP of polariser P2 to $\boldsymbol{S}_a = [0;1;0]$, which results in the following alternative version of Eq. (22),

$$I_x = \bar{I}_y(1+\cos\theta_2)/2. \tag{27}$$

For the second dataset, instead, we set the input SOP at position (1) is restored to $\boldsymbol{S}_1 = [0;-1;0]$ and the analyser SOP is set to $\boldsymbol{S}_a = [0;1;0]$. As a result, Eq. (24) needs to be replaced by

$$I_y = \bar{I}_y(1-\sin\theta_1\sin\theta_2)/2. \tag{28}$$

Using the mapping identified for SLM1, one can fix the pixel value $g_1$ such that $\theta_1 = \pi/2$ and eliminate the dependency on $\theta_1$. Subsequently, the second dataset can be collected. An estimate of $\theta_2$ modulo $2\pi$ is finally obtained from

$$\hat{\theta}_2 = \arctan2(1 - 2I_y/\bar{I}_y, 2I_x/\bar{I}_x - 1). \tag{29}$$

# Supplementary Note 3:
# Correction of SOP and phase in Method A

In Method A we use MM polarimetry [12] to determine the MM corresponding to the aberrating object at position (6) in Supplementary Figure 1. We temporarily mount the object in a separate setup and obtain an estimate of the MM $M$ for each point in an aperture plane, following the method outlined in [12]. The data so obtained for a GRIN lens [12] is reported in Supplementary Figure 4(a) as an example. Afterwards, we convert $M$ into a Jones matrix $J$, as outlined in detail in Supplementary Note 1.3. Using the singular value decomposition (SVD) and the polar decomposition [16, 17], matrix $J$ can be factored into $re^{i\gamma}UH$, where $U$ is special unitary and $H$ is Hermitian. Note that throughout this paper, we always assume that $J$ is homogeneous in the sense outlined in [16]. This is by no means restrictive in our case, as this condition was found to be verified for all experimental data that we collected throughout this study. Factor $U$ corresponds to $\mathrm{SU}2(\boldsymbol{Q}, \delta)$ in Eq. (13). The parameters $\boldsymbol{Q}$ and $\delta$ can be extracted from the eigenvalues $\mu_1$ and $\mu_2$, and eigenvectors $\boldsymbol{q}_1$ and $\boldsymbol{q}_2$ of $U$. In more detail, by applying Eq. (5) to $\boldsymbol{q}_1$ and $\boldsymbol{q}_2$ we obtain respectively $\boldsymbol{Q}$ and $-\boldsymbol{Q}$. The retardance $\delta$ can be obtained as $\delta = 2\arg(\mu_1) = -2\arg(\mu_2)$. Supplementary Figure 4(b)–(c) show the results of applying the analysis outlined above to the data recorded in Supplementary Figure 4(a). After computing the axis $\boldsymbol{Q}$ for each point in the aperture, the spherical angles $\Psi$ and $X$ in the Poincaré sphere can be extracted by inverting the relationship $\boldsymbol{Q} = [\cos(2X)\cos(2\Psi); \cos(2X)\sin(2\Psi); \sin(2X)]$. From $\Psi$ and $X$, the parameters of the polarisation ellipses in the lateral plane can be identified [7] in order to plot the orientation of the extra-ordinary axes across the aperture, as shown in Supplementary Figure 4(b). The corresponding profile of the retardance $\delta$ is reported in Supplementary Figure 4(c). Note that phase unwrapping [14] may be necessary to ensure that the profile of $\delta$ is convex as expected for a GRIN lens [12].

In the absence of the two SLMs and half-wave plate, the aberrating object would change the initial SOP $\boldsymbol{S}_1$ into an aberrated one $\boldsymbol{S}_a$, see Supplementary Figure 5(a). In terms of Jones vectors, this change of SOP is modelled by $\boldsymbol{j}_a = \mathrm{SU}2(\boldsymbol{Q}, \delta)\boldsymbol{j}_1$. The corresponding Stokes vectors $\boldsymbol{S}_1$ and $\boldsymbol{S}_a$ identifying the SOPs can be obtained by applying Eq. (5) to $\boldsymbol{j}_1$ and $\boldsymbol{j}_a$, respectively. Our aim is to apply a pre-correction with the two SLMs such that the final SOP at position (6) still coincides with $\boldsymbol{S}_1$. Equivalently, in terms of Jones vectors, $\boldsymbol{j}_6$ should be equal to $\boldsymbol{j}_1$ up to a global phase. For this purpose, we consider the SOP reached starting from $\boldsymbol{S}_1$ but taking the opposite rotation along the Poincaré sphere, i.e., $\boldsymbol{j}_c = \mathrm{SU}2(\boldsymbol{Q}, -\delta)\boldsymbol{j}_1$, see Supplementary Figure 5(b). The Stokes vector $\boldsymbol{S}_c$ corresponding to $\boldsymbol{j}_c$ is found by applying Eq. (5) to $\boldsymbol{j}_c$. At this point, we can apply Eq. (17) to $\boldsymbol{S}_c$, thus obtaining the necessary values of the retardances $\theta_1$ and $\theta_2$ for the two SLMs. An example of the correction is shown in Supplementary Figure 5(c). First the initial SOP $\boldsymbol{S}_1$ is changed into $\boldsymbol{S}_c$ by the two SLMs and the half-wave plate. This is depicted in the light blue path travelled on the Poincaré sphere in Supplementary Figure 5(c). Afterwards, the beam propagates through the aberrating object, which results in reverting back to the initial SOP $\boldsymbol{S}_1$ – see the yellow path in Supplementary Figure 5(c). Note that, in general, the yellow and blue paths do not coincide. As a result, the overall change in the SOP between position (1) and (6) is a rotation along $\boldsymbol{S}_1$ by an angle $\xi$ – see Eq. (14). This corresponds to a matrix $\mathrm{SU}2(\boldsymbol{S}_1, \xi)$ for which $\boldsymbol{j}_1$ is an eigenvector. For Method A we require a fully-fledged calibration of the SLMs, as described previously in Supplementary Note 2. The DM was calibrated using an interferometer, see optional component I in Supplementary Figure 1(b), as described in detail elsewhere [18].

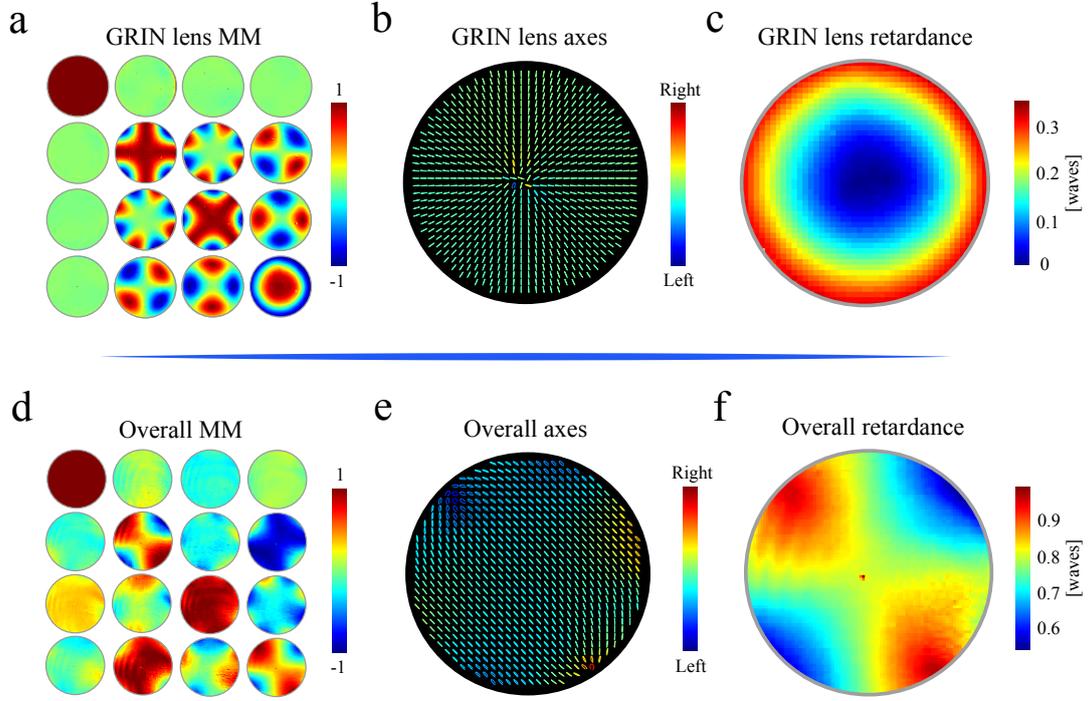

**Supplementary Figure 4:** Correction of a GRIN lens SOP aberration. (a) Mueller matrix data obtained by measuring a cross-section of a GRIN lens using an external MM polarimetry setup described in [12]; (b)–(c) Distribution of the extra-ordinary axes and the retardance $\delta$ of the GRIN lens. Both parameters were extracted from the data reported in (a) using the procedure outlined in Supplementary Note 3. The parameters of the polarisation ellipses are retrieved [7] from the distribution of axis $Q$ across the aperture. The ellipses are then plotted in (b) to visualise the orientation of the extra-ordinary axes. Note that the plot in (b) is different from the plots in Supplementary Figure 2(b)–(c), where the SOP is visualised instead of the axes; (d) Mueller matrix data obtained after correction of the GRIN lens SOP aberration. This measurement is taken between position (1) and position (6) in Supplementary Figure 1, and thus comprises both the pre-correction applied with the two SLMs and the aberration due to the GRIN lens. As can be seen by comparing with (a), element (3,3) has become approximately uniform, which validates the correction for the fixed input SOP $S_1 = [0; -1; 0]$; (e)–(f) Distribution of the axes and retardance after correction of the GRIN lens SOP aberration. Both parameters were extracted from the data reported in (d) As can be seen by comparing (e) with (b), the axes have become approximately linearly polarised along $-45°$, for which the fixed input Jones vector $j_1 = [1; -1]/\sqrt{2}$ is an eigenvector. The retardance shown in (f) essentially becomes a scalar aberration, which can be corrected with the DM.

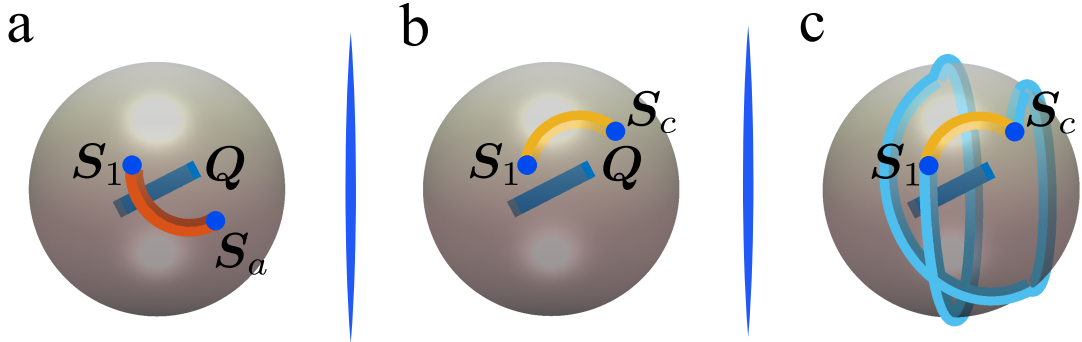

**Supplementary Figure 5:** Example showing the correction of the SOP on the Poincaré sphere. (a) An object aberrating the SOP is represented by the Jones matrix $SU2(\boldsymbol{Q}, \delta)$ and can be visualised as a rotation by an angle $\delta$ about the axis $\boldsymbol{Q}$; (b) The optimal pre-correction $\boldsymbol{S}_c$ is found by starting from $\boldsymbol{S}_1$ and by applying the opposite rotation $SU2(\boldsymbol{Q}, -\delta)$; (c) Correction of the SOP. Starting from $\boldsymbol{S}_1$, the two SLMs and half-wave plate change the SOP into $\boldsymbol{S}_c$. This is depicted as a set of rotations along the light blue paths. Subsequently, the aberrating object applies the rotation $SU2(\boldsymbol{Q}, \delta)$ depicted as the yellow path, which restores the SOP to the initial state $\boldsymbol{S}_1$.

The results after applying the SOP correction are reported in Supplementary Figure 4(d)–(f). Supplementary Figure 4(d) shows the MM data measured from the input position at (1) and the output position at (6) – see Supplementary Figure 1. This comprises both the pre-correction applied with the two SLMs as well as the effect of the GRIN lens SOP aberration. As can be seen by comparing this figure to Supplementary Figure 4(a), element (3, 3) has become approximately uniform, which validates the correction for the fixed input SOP $\boldsymbol{S}_1 = [0; -1; 0]$ as desired. Note that a side effect of the correction is that the modulation for other input SOPs has also changed. This is expected, as the two SLMs module can only correct for a fixed input SOP [9]. In Supplementary Figure 4(e), analysis of the MM data as previously described also reveals that the distribution of the axes has become approximately uniform – compare with Supplementary Figure 4(b). The axes are approximately linearly polarised along $-45°$, for which the fixed input Jones vector $\boldsymbol{j}_1 = [1; -1]/\sqrt{2}$ is an eigenvector. Finally, we note that, since $\boldsymbol{j}_1 = [1; -1]/\sqrt{2}$, the overall retardance depicted in Supplementary Figure 4(f) results in a scalar phase aberration, which can be corrected with the DM.

After applying the SOP correction as outlined above, we correct the scalar phase by employing conventional wavefront sensorless AO [19]. This iterative procedure, which is described in detail elsewhere [8, 19, 20], optimises the shape of the DM by repeatedly measuring the FID with camera C2 in Supplementary Figure 1. This concludes the vectorial adaptive optics (V-AO) aberration correction, whereby both the SOP and scalar phase have been compensated using Method A.

## Supplementary Note 4:
## Sensor-based and sensorless vectorial adaptive optics

Conventional adaptive optics (scalar phase AO) is often applied following one of two different strategies. The first approach, which we refer to as sensor-based, relies on a wavefront sensor (WFS) such as a Shack-Hartmann WFS [21, 22], which is deployed at a plane conjugate to the pupil of the optical system. An alternative approach, which we call sensorless [8, 19, 20, 23], does not rely on a WFS. Instead, aberration correction is achieved iteratively, by optimising an image-quality metric. In this strategy, no measurements of the wavefront are performed at the pupil plane. Only the images obtained with the optical system or the FID such as in our case are used to indirectly determine and correct the aberration. Both approaches are listed within the phase sub-tree in Supplementary Figure 6.

In contrast to phase AO, V-AO requires a method to determine the full vectorial state of the aberration, which comprises both the phase and SOP. The SOP can be measured via a Mueller matrix (MM) polarimeter, which is depicted in Supplementary Figure 7. The polarimeter consists of two parts, namely a polarisation state generator (PSG) and polarisation state analyser (PSA) [24]. By using a polarimeter, a strategy that is analogous to the sensor-based strategy for phase AO can be implemented. This approach was adopted with Method A in Supplementary Note 3.

Strategies for V-AO that are analogous to sensorless phase AO can also be devised. These MM polarimeter-free approaches are shown under the sensorless node for polarisation in Supplementary Figure 6. The first one, highlighted in blue, concerns the quasi-sensorless method outlined in detailed in Supplementary Note 5, whereby only a subset of the measurements necessary for full MM polarimetry are taken at a plane conjugate to the pupil of the system. Since the MM polarimetry measurements are incomplete, an iterative approach similar to sensorless phase AO is implemented to achieve the SOP correction. A second kind of sensorless approach, highlighted in pink in Supplementary Note 5, does not rely on any measurements at the pupil plane at all. Instead, only measurements at the focus are acquired and an iterative optimisation of an image-quality metric is performed. In this case, some prior knowledge is assumed about the SOP aberration. This approach is implemented in Supplementary Note 7.

We remark that in this paper we address both the issues of correcting the vectorial aberration as well as the problem of enhancing the focal spot via V-AO modulation. As far as Method A and Method B are concerned, see Supplementary Note 3 and Supplementary Note 5, we are able to achieve both correction of the vectorial aberration and improvement of the focal spot. For Method C instead, see Supplementary Note 7, our primary concern is enhancing the focal spot and improving the uniformity of the SOP.

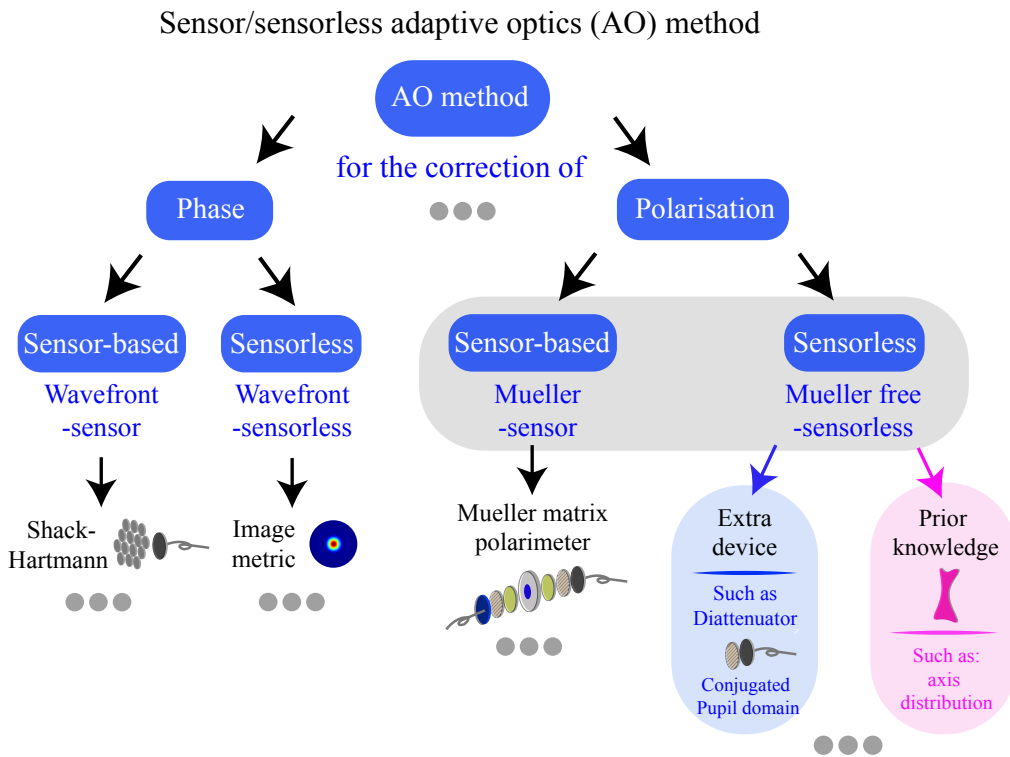

**Supplementary Figure 6:** Classification of the sensor-based and sensorless strategies for conventional phase-only AO and for V-AO.

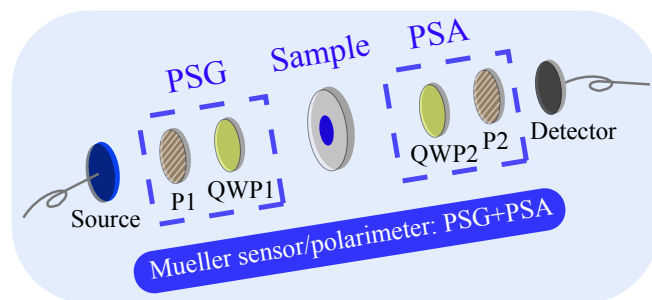

**Supplementary Figure 7:** Layout of a MM polarimeter. The PSG and PSA are highlighted in blue dotted boxes. P polariser; QWP rotating quarter-wave plate.

# Supplementary Note 5:
# Correction of SOP and phase in Method B

In Method B we assume that the parameters $Q$ and $\delta$ of SU2$(Q,\delta)$ in Eq. (13) are unknown and we do not employ a MM polarimeter for their estimation, contrary to what was done in Supplementary Note 3. Instead, correction of the SOP is achieved using a set of intensity measurements in the pupil plane, which is obtained using polariser P2 and camera C1 in Supplementary Figure 1.

By examining Eq. (17), one can see that $S_4$ can span every point in the Poincaré sphere, even if the intervals of $\theta_1$ and $\theta_2$ are restricted to $\theta_1 \in [0,\pi]$ and $\theta_2 \in [0,2\pi]$, respectively. For our purposes it is therefore only sufficient for the first SLM to induce up to half a period of retardance. Next, we introduce a further simplification with respect to Supplementary Note 3, by not requiring a fully-fledged calibration of the SLMs, as outlined in Supplementary Note 2. In more detail, we only require knowledge of the endpoints of the pixel value intervals $g_1 \in [l_1, h_1]$ and $g_2 \in [l_2, h_2]$, respectively corresponding to $\theta_1 \in [0,\pi]$ and $\theta_2 \in [0,2\pi]$. This requirement allows for a significant reduction in the effort to characterise the SLMs, since we do not need to identify in detail the relationship between $g_1$ and $\theta_1$, as well as that between $g_2$ and $\theta_2$. We therefore proceeded to establish the problem of correcting the SOP directly in terms of $g_1$ and $g_2$, instead of $\theta_1$ and $\theta_2$. Considering a single point in the pupil measured with camera C1, the intensity as a function of $g_1$ and $g_2$ peaks when the SOP at position (6), i.e., $S_6$ is aligned with the SOP transmitted by the polariser P2. As a result, the problem of correcting the SOP can be solved by maximising the intensity measured by camera C1 as a function of $g_1$ and $g_2$.

We adopt the following strategy to solve the intensity maximisation problem outlined above. First a coarse grid of sampling points $\mathcal{G}_c$ is created within the Cartesian product $[l_1, h_1] \times [l_2, h_2]$, i.e., $\mathcal{G}_c \in [l_1, h_1] \times [l_2, h_2]$. For each tuple $(g_1, g_2)$ in $\mathcal{G}_c$, we assign $g_1$ to all pixels of the first SLM and $g_2$ to all pixels of the second SLM. After that we record an image with camera C1. The same steps are repeated until an image has been recorded for each tuple in $\mathcal{G}_c$. At this point, we have all the information to estimate the optimal pixel values for each point within the pupil. In more detail, for each point in the pupil, we evaluate the interpolation of $\mathcal{G}_c$ over the full Cartesian product $[l_1, h_1] \times [l_2, h_2]$ – the number of tuples in the product is finite, since pixel values are integral. Finally, the optimal values for $g_1$ and $g_2$ are identified by selecting the maximum value from the interpolation of $\mathcal{G}_c$. Note that this process can be performed in parallel for each pixel within the aperture disk.

In our experimental setup, the intervals for the pixel values were $g_1 \in [8, 148]$ and $g_2 \in [44, 156]$. We sampled the two intervals with 8 points for each, resulting in the coarse grid

$$\mathcal{G}_c = [8, 28, 48, 68, 88, 108, 128, 148] \times [44, 60, 76, 92, 108, 124, 140, 156]. \qquad (30)$$

For each tuple in $\mathcal{G}_c$, we first applied the corresponding pixel values $g_1$ to all pixels of SLM1 and $g_2$ to all pixels of SLM2. Second, we recorded the corresponding image with camera C1. Overall, this results in a set of 64 images. After collecting this data, the procedure to identify the optimal pixel value is performed in parallel for each pixel within the pupil. We plot the intermediate results for a single pixel in Supplementary Figure 8(a) and Supplementary Figure 8(b). In (a), the measured intensity for the pixel is plotted versus the coarse grid $\mathcal{G}_c$. In (b), the sampled values from (a) are interpolated and the maximum intensity is selected, thus identifying the optimal values for $g_1$ and $g_2$ for this pixel. Once the optimal values for each pixel have been identified, one can construct the complete pattern for SLM1 from the values of $g_1$, see Supplementary Figure 8(c). Similarly, the complete pattern for SLM2 is constructed from the optimal values of $g_2$, see Supplementary Figure 8(d). We remark that the patterns computed in this way may not be directly applicable to the two SLMs, since the number of pixels within the aperture recorded by camera C1 may not match the number of pixels encompassed within the aperture for SLM1 and SLM2. This difficulty can be easily overcome by using interpolation. Finally, we empirically applied

Gaussian filtering with a standard deviation of 8 pixels to smoothen the patterns. Once the patterns are applied with the two SLMs this concludes the procedure for the correction of the SOP aberration. The subsequent correction of the scalar phase is performed as outlined previously in Supplementary Note 3 for Method A. A flow chart summarising the main steps for Method B is found in Supplementary Figure 9.

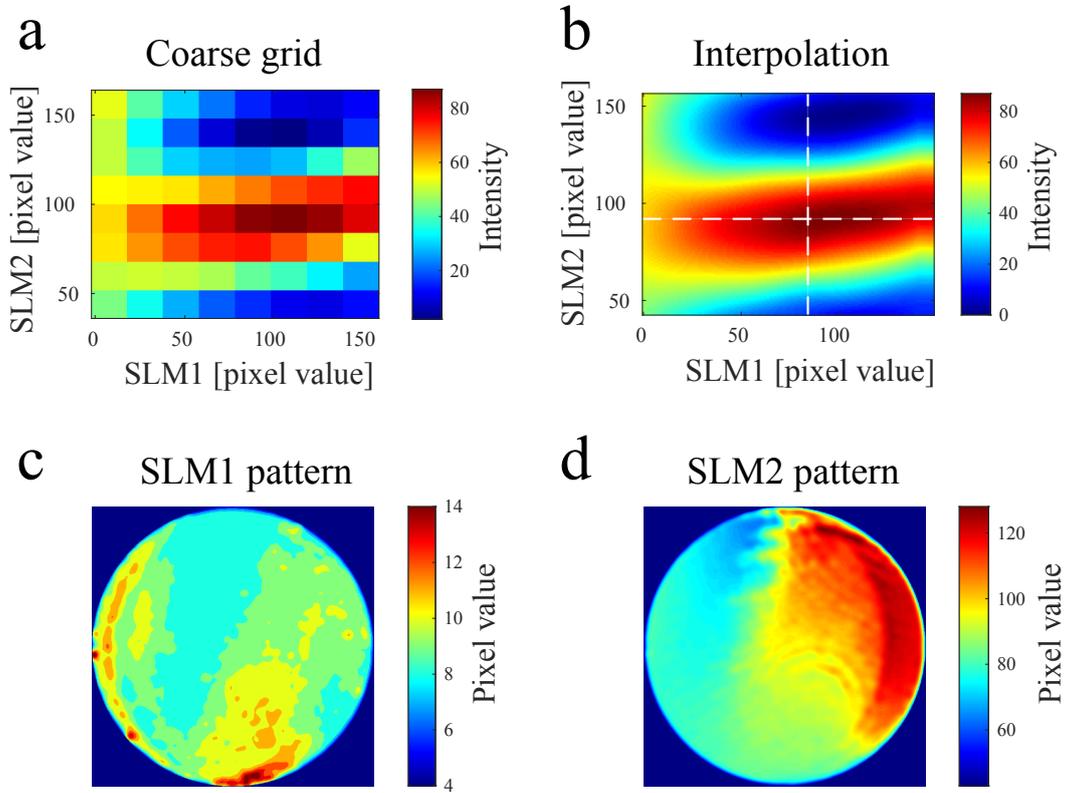

**Supplementary Figure 8:** Intermediate steps performed during the correction of the SOP for Method B. A coarse grid of pixel values tuples is initially created, see Eq. (30). For each tuple $(g_1, g_2)$, $g_1$ is applied to all pixels of SLM1, $g_2$ is applied to all pixels of SLM2, and an image is recorded with camera C1, as outlined in Supplementary Note 5. For each pixel in the aperture, the intensity sampled from the recorded images and its interpolation are evaluated, see (a) and (b) respectively, and the maximum intensity is identified in (b). From the pixel values $g_1$ and $g_2$ corresponding to the maximum interpolated intensity, the correction patterns for SLM1 and SLM2 are computed, as shown in (c) and (d), respectively.

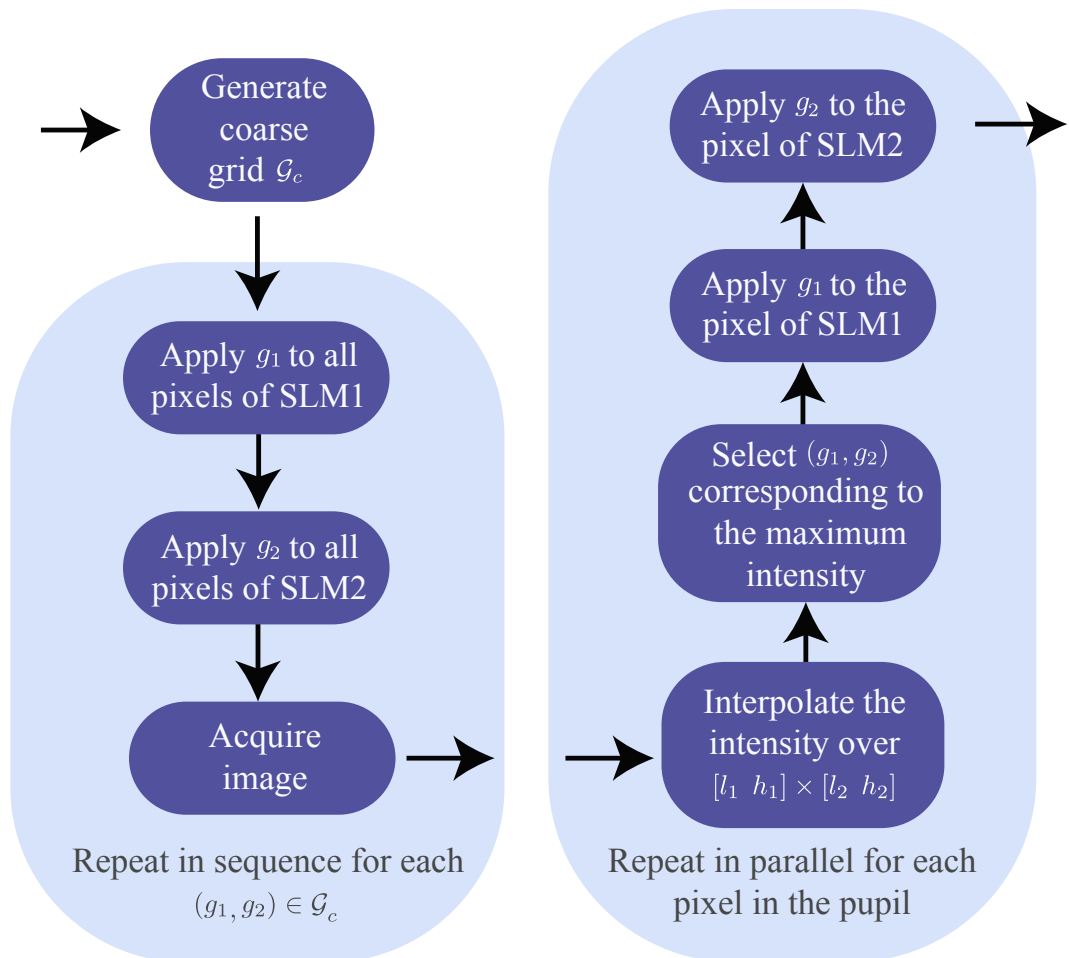

**Supplementary Figure 9:** Flow chart summarising the main steps for Method B outlined in Supplementary Note 5. The final correction of the phase using a sensorless method [19] is not depicted.

# Supplementary Note 6:

# Analysis of vectorial aberrations arising from mirrors

In the main article we demonstrate aberration correction for a sequence of protected silver mirrors, which are commonly used in many optical systems. This illustrates the importance of spatially variant V-AO correction. In this note, we analyse in more detail the polarisation aberrations arising from such mirrors and other commonly used optical components such as beam splitters and bandpass filters. We start by measuring the aberration due to a single protected silver mirror when this is reflecting a collimated beam at a variable angle of incidence $\beta$. The results are shown on the left in Supplementary Figure 10(a), where one can see that the induced retardance increases as a function of $\beta$. Note that the measured diattenuation is approximately invariant of $\beta$ and can therefore be neglected as done in this paper. The aberration arising from pairs of mirrors oriented at 45° is also reported on the right in Supplementary Figure 10(a). This layout is commonly employed in optical microscopes and other optical equipment.

    In Supplementary Figure 10(b), we report the same measurements for other commonly used elements, namely cube, plate, and pellicle beam splitters, and a bandpass filter (633 nm transmission). Measurements are performed both in a transmission and in a reflection ($\beta = 45°$) geometry, which are denoted by T and R on the abscissa. The vectorial aberrations arise due to coatings (thin film), Fresnel's effects, and other factors – see [1, 25–32] for further details. The presence of such aberrations further highlights the usefulness of our V-AO aberration correction techniques.

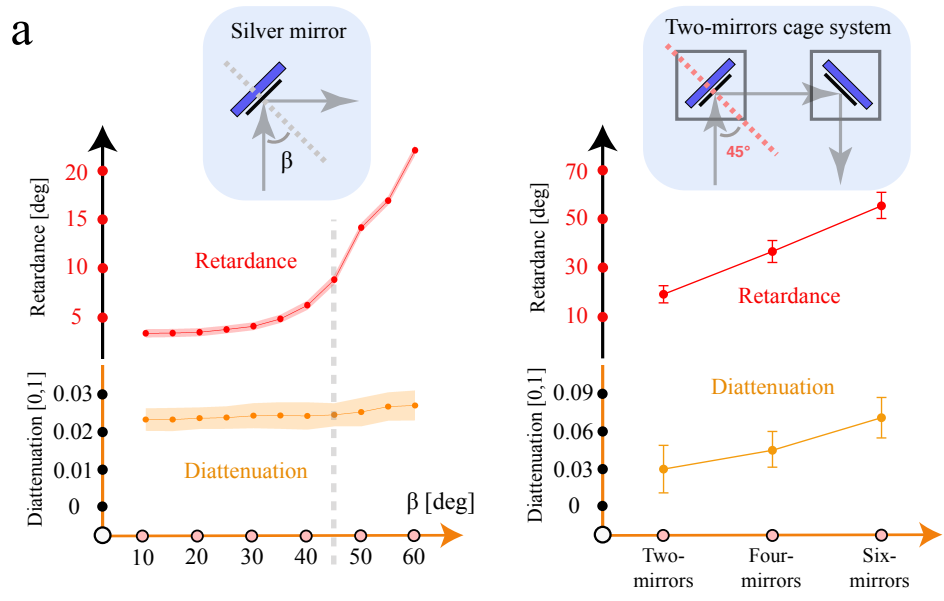
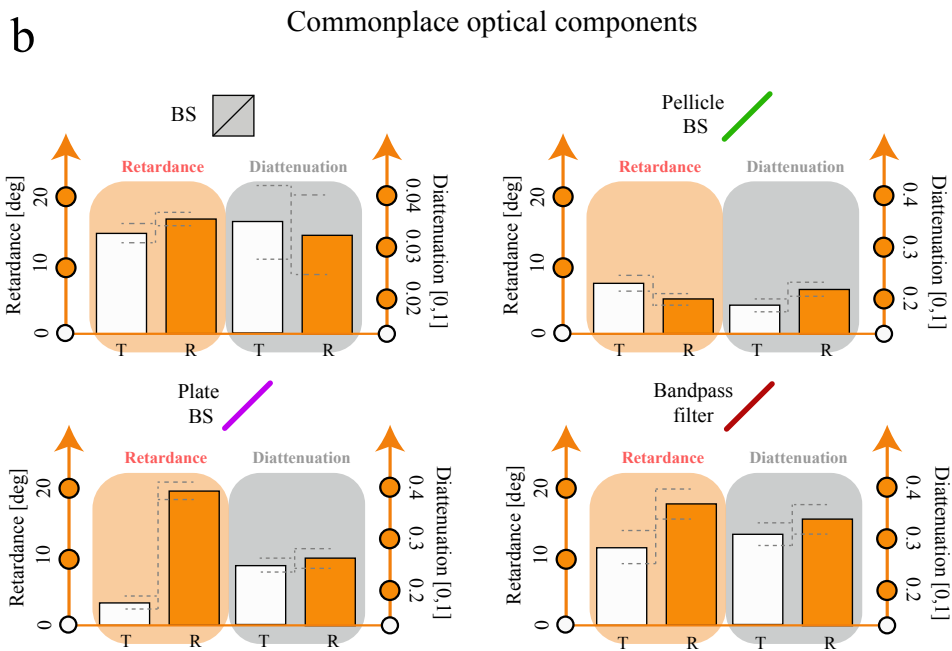

**Supplementary Figure 10:** Vectorial aberrations arising from common optical elements. (a) On the left, the retardance and diattenuation measured due to a silver mirror oriented at different angles of incidence $\beta$. On the right, the same measurements using one, two, and three pairs of mirrors oriented at 45°; (b) Vectorial aberrations obtained from cube, plate, and pellicle beam splitters, and a bandpass filter (633 nm transmission). The measurements are performed in transmission and reflection denoted by T and R in the abscissa.

# Supplementary Note 7:

# Correction of SOP and phase in Method C

In Method C we investigate applying iterative, image-based methods to improve the FID by modulating both the SOP and the phase. These methods are inspired by conventional wavefront sensorless algorithms, which are employed to correct scalar phase aberrations in microscopy applications [8]. For Method C, therefore, we do not rely on measurements in the pupil plane, in contrast to Method B in Supplementary Note 5. Instead, only measurements of the FID are available, obtained with camera C2 – see Supplementary Figure 1. For Method C we employ a fully-fledged calibration of the SLMs, as described previously in Supplementary Note 2.

As seen from Eq. (13), Eq. (14), and Eq. (15), modulation of the SOP and the scalar phase are deeply intertwined. Therefore, the task of simultaneously optimising both SOP and phase using image-based methods appears extremely challenging. Henceforth, to make the problem more tractable, we introduce some prior knowledge about the aberrating object at position (6). In more detail, we assume that the distribution of $Q$ over the pupil disk is known a priori. Nonetheless, the retardance $\delta$ remains unknown and must still be determined via the iterative optimisation.

## 7.1 Scenario 1

In the first scenario we assume that the object can be modelled as a positive uniaxial material, see Eq. (6). We further assume that the object does not induce any scalar phase aberration due to OPL differences. As a result, we have that $\phi = \delta$ and that Eq. (13) can be simplified as

$$
\begin{aligned}
j_6 = {} & \exp\{i[\psi + \delta + (\theta_1 + \theta_2)/2]\} \\
& \times \text{SU2}(Q, \delta) \\
& \times \text{SU2}(H, \pi)\,\text{SU2}(H, \theta_2) \\
& \times \text{SU2}(W, \pi) \\
& \times \text{SU2}(H, \pi)\,\text{SU2}(H, \theta_1) \\
& \times j_1.
\end{aligned}
\tag{31}
$$

Note that Eq. (31) accounts for a single point in the pupil only. Nevertheless, one must determine values for $\theta_1$, $\theta_2$, and $\psi$ throughout the whole pupil in order to apply a single correction. A point-wise optimisation of $\theta_1$, $\theta_2$, and $\psi$ in the pupil using a sensorless method is therefore impractical, due to the large number of degrees of freedom (DOF) involved. For this reason, we employ Zernike polynomials, as defined in [33], in order to reduce the number of DOFs to a more tractable level. We use a vector $\boldsymbol{\alpha}$ of $N_z$ Zernike coefficients to describe the spatially variant retardance profile across the pupil, i.e.,

$$
\hat{\delta}(\rho, \varphi) = \sum_{i=1}^{N_z} \alpha_i \mathcal{Z}_i(\rho, \varphi), \tag{32}
$$

where $\alpha_i$ are the elements of $\boldsymbol{\alpha}$, and $\rho$ and $\varphi$ are the polar coordinates in the pupil disk. $\mathcal{Z}_i$ is the $i$-th Zernike polynomial which is defined elsewhere [33]. Vector $\boldsymbol{\alpha}$ is the independent variable to be optimised by the sensorless algorithm throughout the iterations. For each point $(\rho, \varphi)$ in the pupil, we can use Eq. (32) to compute an estimate $\hat{\delta}$ of the true retardance $\delta$, which allows us to compute $\hat{U} = \text{SU2}(Q, \hat{\delta})$, since $Q$ is known a priori. Subsequently, we follow Supplementary Note 3, by calculating the pre-corrected Jones vector

$$
\hat{j}_c = \text{SU2}(Q, -\hat{\delta})\,j_1, \tag{33}
$$

and its corresponding Stokes vector

$$\hat{\boldsymbol{S}}_c = \begin{bmatrix} \hat{\boldsymbol{j}}_c^\dagger \sigma_1 \hat{\boldsymbol{j}}_c \\ \hat{\boldsymbol{j}}_c^\dagger \sigma_2 \hat{\boldsymbol{j}}_c \\ \hat{\boldsymbol{j}}_c^\dagger \sigma_3 \hat{\boldsymbol{j}}_c \end{bmatrix}, \qquad (34)$$

retrieved with Eq. (5). Finally, we apply Eq. (17) to $\hat{\boldsymbol{S}}_c$, thus obtaining the corresponding values of the retardances $\theta_1$ and $\theta_2$ that must be applied with the two SLMs. These latter operations can be performed in parallel for each point in the pupil. Note that, if the condition $\hat{\delta} = \delta$ holds true then this procedure ensures that the SOP is corrected as desired.

It is important to realise that correction of the SOP alone as outlined above is not a sufficient condition to obtain a diffraction-limited FID. In fact, even assuming that $\hat{\delta} = \delta$, correcting the SOP can induce a non-zero scalar phase across the pupil. This phase is due to the contributions of $\theta_1$ and $\theta_2$ in Eq. (31), in addition to the contribution due to $\xi$ in Eq. (14). The induced scalar phase is a serious impediment to the sensorless algorithm, which can undermine the correction of the SOP. For example, consider the situation whereby the non-zero phase renders the FID more distorted with respect to its state before the SOP correction was applied. In such an instance, the sensorless algorithm fails to select the second state as the more optimal one, since the FID was degraded further. Instead, the first state comprising the SOP aberration but exhibiting a sharper FID is identified as the more optimal one and the SOP aberration is not corrected.

To try to overcome this critical issue one can compensate the scalar phase with the DM just after applying the SOP modulation but before recording the FID used by the sensorless algorithm. A possible strategy is to use the values of $\theta_1$, $\theta_2$, and $\hat{\delta}$ computed above and Eq. (31) to try to predict the Jones vector at position (6), i.e., $\hat{\boldsymbol{j}}_6$. Subsequently, the phase difference between $\boldsymbol{j}_1$ and $\hat{\boldsymbol{j}}_6$ can be computed using the Pancharatnam connection [6, 10], which we define as

$$\Pi(\boldsymbol{a}, \boldsymbol{b}) = \begin{cases} \arg(\boldsymbol{a}^\dagger \boldsymbol{b}) & \boldsymbol{a}^\dagger \boldsymbol{b} \neq 0 \\ 0 & \boldsymbol{a}^\dagger \boldsymbol{b} = 0 \end{cases}. \qquad (35)$$

This phase can be finally cancelled using the DM by letting $\psi = -\Pi(\boldsymbol{j}_1, \hat{\boldsymbol{j}}_6)$. Now, if one has that $\hat{\delta} = \delta$, then a diffraction-limited FID is obtained as desired, since the phase induced by the SOP modulation has been suppressed. During the iterations of the sensorless algorithm one obviously has that $\hat{\delta} \neq \delta$, as otherwise the ideal correction would already have been attained. In such a case, the SOP of $\boldsymbol{j}_1$ and $\hat{\boldsymbol{j}}_6$ do not match, i.e., $\boldsymbol{S}_1 \neq \hat{\boldsymbol{S}}_6$, and phase comparison between $\boldsymbol{j}_1$ and $\hat{\boldsymbol{j}}_6$ should be performed using the Pancharatnam connection given in Eq. (35). We emphasise that the algorithm presented so far is a heuristic approach that leads successfully to correction of the SOP uniformity. Our experimental results show that the algorithm yields consistent improvement of the FID.

Below we report some illustrative data obtained during the correction of the SOP induced by a calibration target, which is described in the main article. The distribution of $\boldsymbol{Q}$ along the aperture was obtained using MM polarimetry, as described previously in Supplementary Note 3, and is reported in Supplementary Figure 12. The first 10 Zernike modes were considered for Eq. (32). The procedure begins by optimising the first Zernike mode, i.e., piston. Different amounts of piston, i.e., $\{-1.2, -0.8, -0.4, 0.0, 0.4, 0.8, 1.2\}$, are applied to $\alpha_1$ in Eq. (32), and $\hat{\delta}$ is computed – see the abscissa in Supplementary Figure 13(a) and the first column in Supplementary Figure 14. For each point of $\hat{\delta}$ in Eq. (32), $\hat{\boldsymbol{j}}_c$ and its corresponding Stokes vector $\hat{\boldsymbol{S}}_c$ are computed, from which the values of $\theta_1$, $\theta_2$, and $\psi$ are determined as explained above. The computation of $\theta_1$, $\theta_2$, and $\psi$ is carried out in parallel for each pixel in the pupil. The values so computed are shown in the second to fourth columns in Supplementary Figure 14 for three different amounts of piston marked with letters [A], [B], and [C] in Supplementary Figure 13(a). After the SLMs and DM are updated accordingly to the values of $\theta_1$, $\theta_2$, and $\psi$, the corresponding measurement of the FID is recorded with camera C2 – see the last column of Supplementary Figure 14. For each measured FID, the value of the image quality metric [19] is also evaluated. This allows us to fit a quadratic function through the data points of the metric – see Supplementary Figure 13(a), where the data points of the metric are plotted in blue and the fit is shown in red. Finally, the optimal value of piston is identified – see the dashed vertical bar in Supplementary Figure 13(a). This allows us to determine the optimal $\alpha_1$ in $\boldsymbol{\alpha}$ and concludes the optimisation of piston for $\hat{\delta}$.

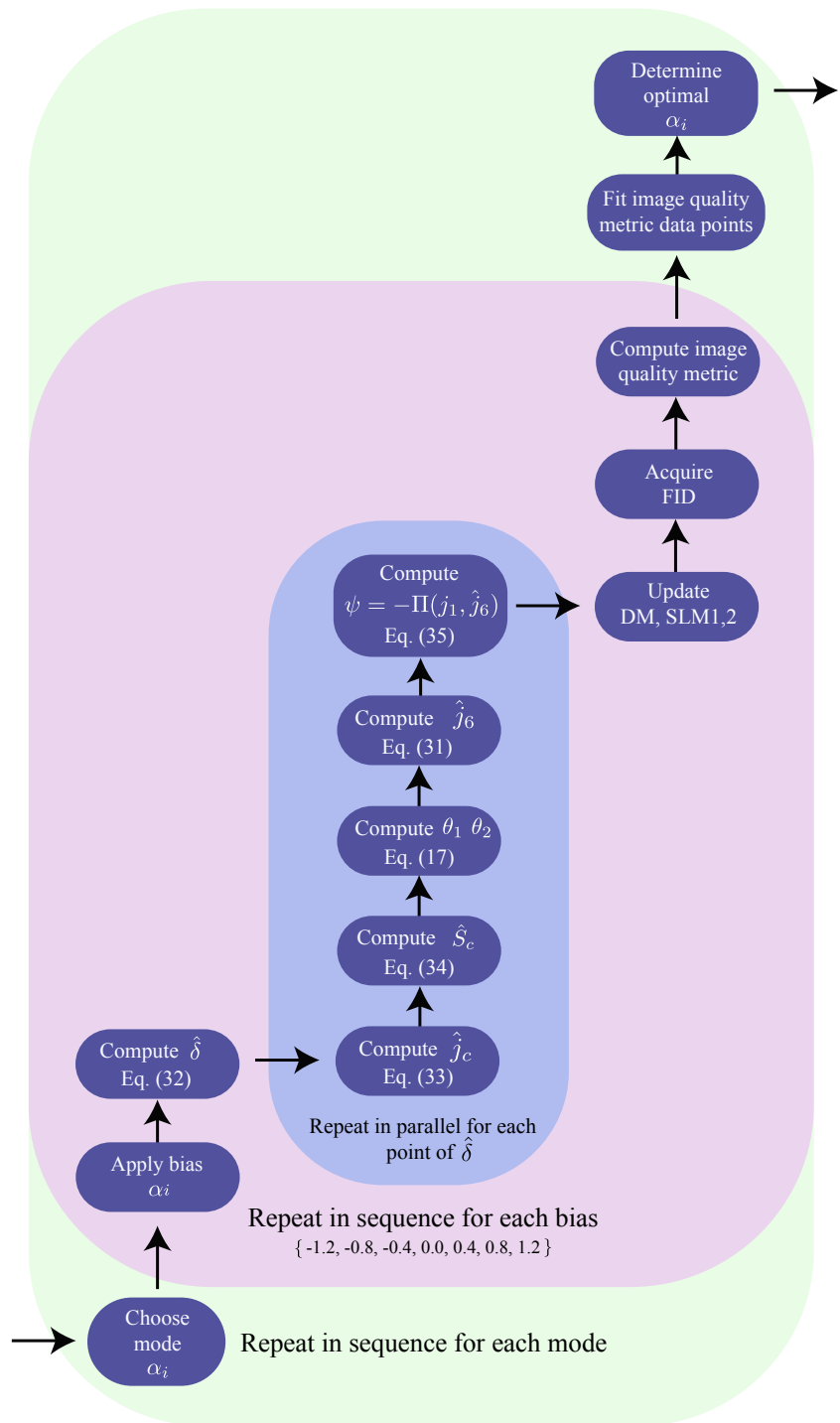

**Supplementary Figure 11:** Flow chart summarising the main steps for Method C outlined in Supplementary Note 7.1.

The procedure explained in the paragraph above is then repeated for the remaining Zernike modes. Supplementary Figure 13(b) and Supplementary Figure 15 show some of the steps taken during the optimisation of the fourth mode, i.e., defocus. After the optimisation is concluded, we applied MM polarimetry to measure the MM between position (1) and position (6) and extracted the parameters of the axes of the combined pre-correction and aberrating target as described previously. The results are shown in Supplementary Figure 12(b).

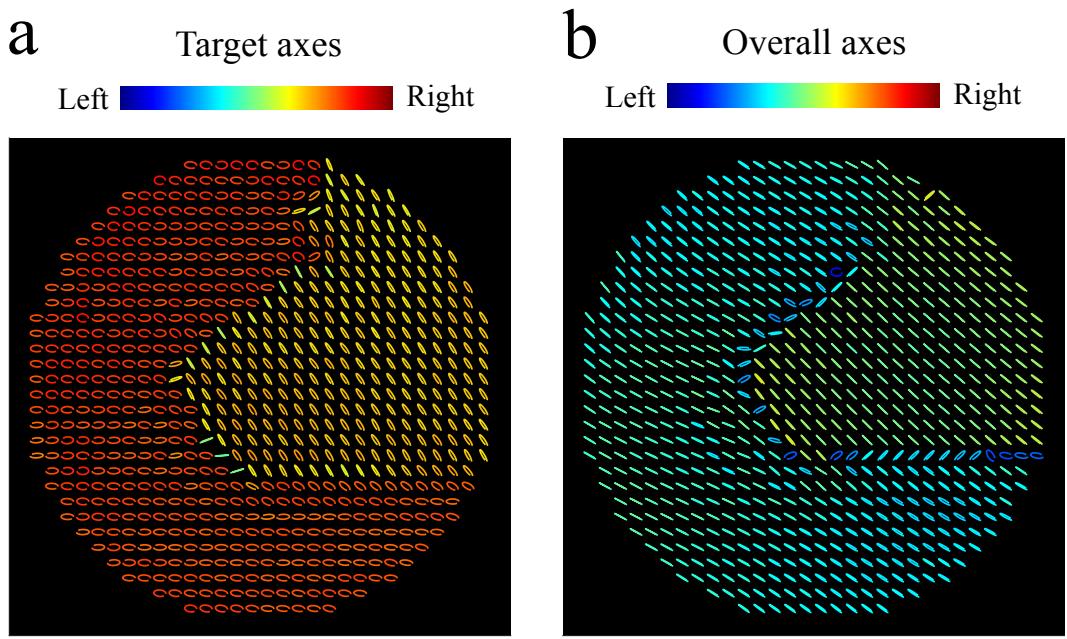

**Supplementary Figure 12:** Correction of the SOP aberration due to a target object. (a) Distribution of the axes of the target. This information is assumed known a priori in Method C. Parameter $\boldsymbol{Q}$ is extracted from MM polarimetry measurements, as outlined in Supplementary Note 3. From the relationship $\boldsymbol{Q} = [\cos(2X)\cos(2\Psi); \cos(2X)\sin(2\Psi); \sin(2X)]$ the parameters of the polarisation ellipse are obtained [7], and the corresponding ellipses are plotted in the aperture disk; (b) Distribution of the axes of the combined pre-correction using the SLMs and the target aberrating object at position (6) after applying Method C, Scenario 1, see Supplementary Note 7.1. The axes are approximately homogeneously linearly polarised, aside from errors at the boundary, showing improvement with respect to (a).

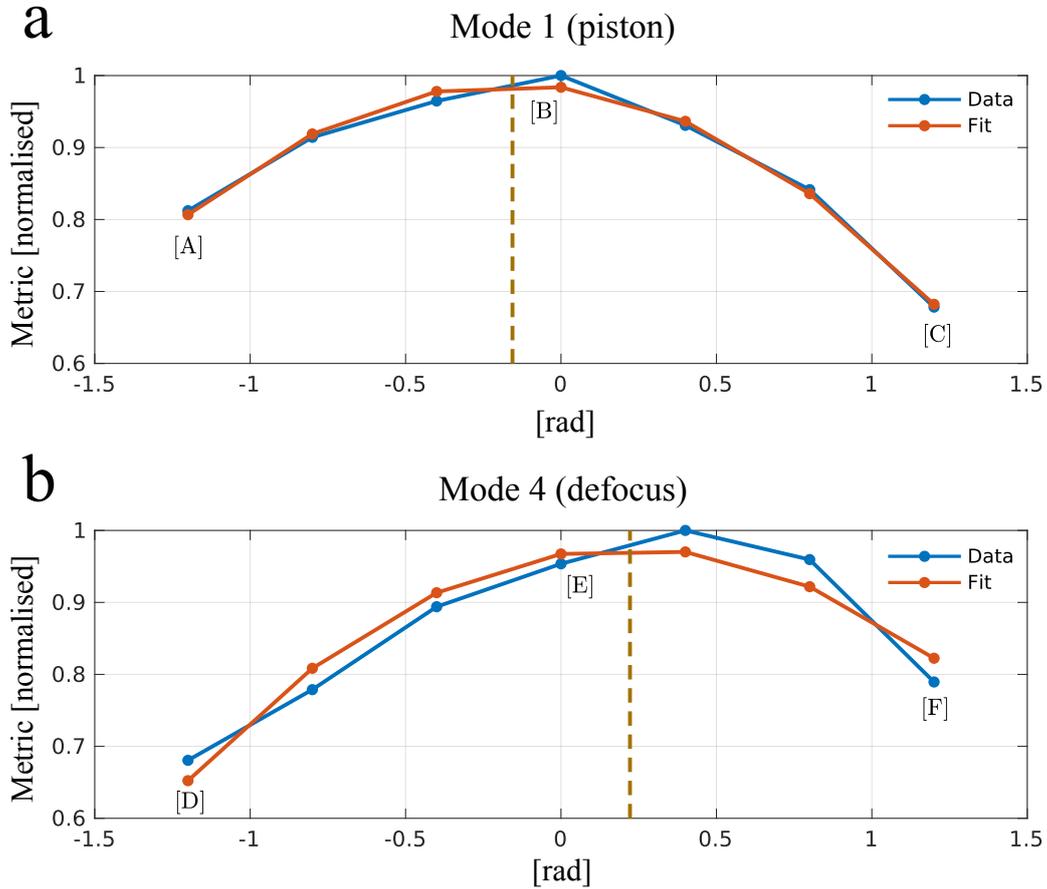

**Supplementary Figure 13:** Example of fits performed for the sensorless algorithm described in Supplementary Note 7.1. (a) Measurements of the metric in blue and fit of the measurements in red for the first Zernike mode $\mathcal{Z}_1$; (b) Same as (a) for the fourth Zernike mode $\mathcal{Z}_4$ [33]; In both (a) and (b), the abscissa denotes the amount of Zernike mode applied. The ordinate denotes the normalised value of the image quality metric [19] determined measuring the corresponding FID. The vertical dashed line denotes the optimal amount of mode determined after the fit. The value of $\hat{\delta}$, $\theta_1$, $\theta_2$, and $\psi$ is annotated at some states with letters [A]–[F], see Supplementary Figure 14 and Supplementary Figure 15. The correction is applied incrementally, so that in (b) the correction for modes 1, 2, and 3 is already applied.

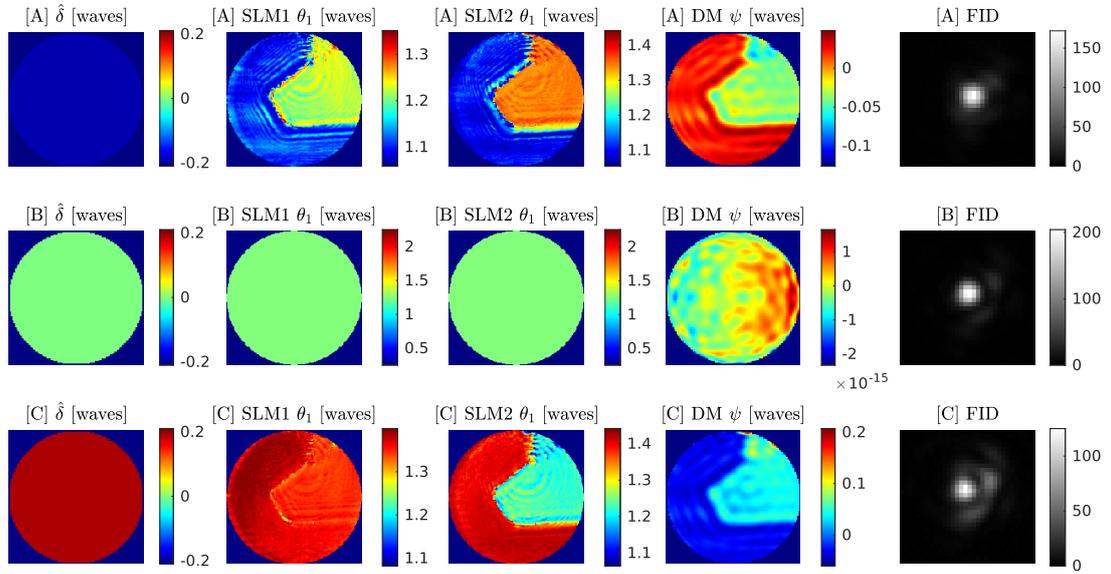

**Supplementary Figure 14:** State of $\hat{\delta}$, $\theta_1$, $\theta_2$, and $\psi$ at different steps annotated by letters [A], [B], and [C] during the correction of mode 1 in Supplementary Figure 13(a).

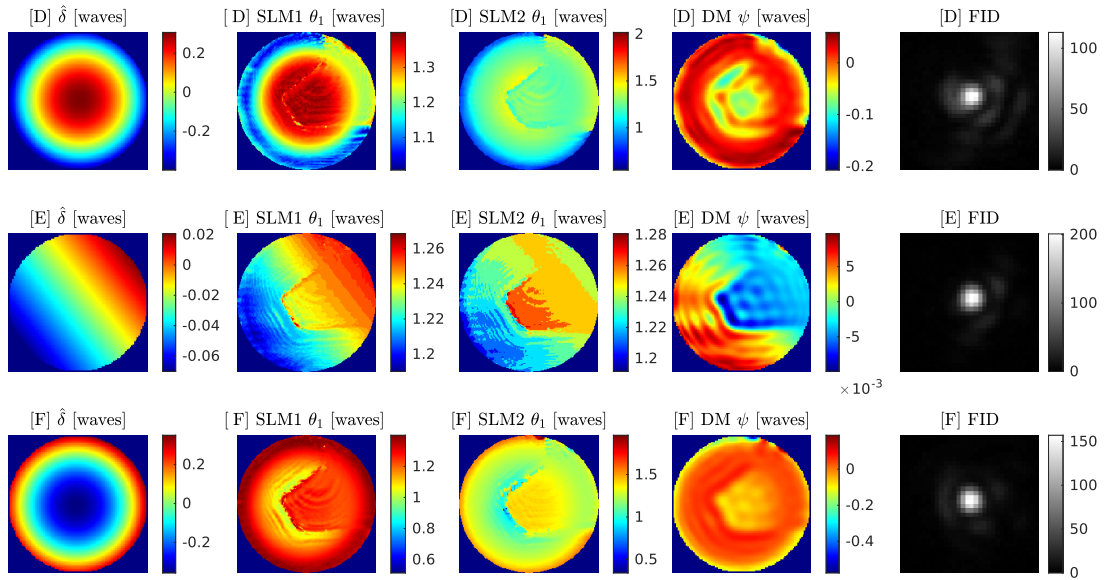

**Supplementary Figure 15:** State of $\hat{\delta}$, $\theta_1$, $\theta_2$, and $\psi$ at different steps annotated by letters [D], [E], and [F] during the correction of mode 4 in Supplementary Figure 13(b).

## 7.2 Scenario 2

In the second scenario we instead consider Eq. (13) and remove the assumption made previously that $\phi = \delta$. In this case, the phase term $\phi$ accounts for the cumulative effects of OPL differences and the geometric phase. We therefore employ a heuristic approach to perform the full V-AO aberration correction in three steps. The first step involves applying conventional sensorless AO [19] to correct most of the aberration due to OPL differences across the pupil. In the second step we apply the procedure described earlier in Supplementary Note 7.1. Finally, in the last step, we apply conventional sensorless AO [19] for a second time. The last phase correction applied in the first step is used as a fixed offset to the phase applied with the DM in the second step. In this way, the phase correction obtained in the first step is used as an initial condition for the phase modulation that occurs in the second step. The same applies for the phase correction determined in the second step and the third step. Our experimental results show consistent improvement of the FID when this three-steps procedure is used.

# Supplementary Note 8:
# Discussion

Throughout this work we have dealt with the correction of the retardance vectorial aberration. However, there exist other types of polarisation aberrations, e.g., depolarisation [1], whereby the degree of polarisation (DOP) is also affected. In this case, the SOP correction can still be conducted via our current approach. On the other hand, polarizance needs to be taken into consideration for further V-AO developments addressing depolarisation.

Another possible extension to consider is the correction of an object comprising multiple layers – this case is relevant for deep tissue imaging, where conventional phase-only AO already plays a crucial role [8]. For V-AO, the situation is significantly more involved, due to issues arising from reciprocity and the lack of the commutative property in the matrix product. For instance, the V-AO aberration experienced when a beam is focussed into a specimen may differ from the one encountered when the beam propagates back in the opposite direction. A typical scenario is depicted in Supplementary Figure 16, where a beam of light is focussed into an object consisting of multiple layers highlighted using different colours in the inset on the right. Each layer can be considered as a homogeneous element such as a pure retarder, depolariser, or diattenuator. The incident beam coming from the PSG would be modulated by the first layer $L_1$, the second one $L_2$, and so forth until layer $L_n$. Instead, upon reflection from the focal spot, modulation occurs in the reverse order before the beam is reflected by the beam splitter and impinges onto the PSA. Using Mueller calculus, this can be modelled as

$$\boldsymbol{S}_{\text{out}} = M_{\text{PSA}} M_{L_1} \cdots M_{L_n} \cdots M_{L_1} M_{\text{PSG}} \boldsymbol{S}_{\text{in}}. \tag{36}$$

We remark that in Eq. (36) we obviously consider full Stokes vectors with four components, contrary to the case for the other supplementary notes. To deal with this focussing arrangement we may introduce two V-AO modules, denoted by A and B in Supplementary Figure 16. Each module is responsible for cancelling the aberration due to the passes through the multi-layer object in the correct order. Further developments would be necessary to apply V-AO in this scenario, perhaps exploiting biomarkers such as nano-probes [34].

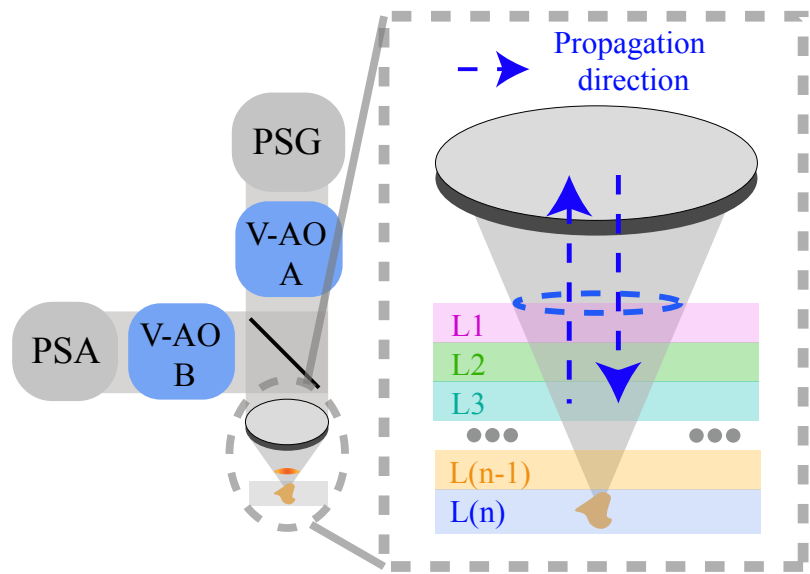

**Supplementary Figure 16:** Vectorial aberrations in a multi-layer object. A focussing arrangement is shown on the left where two V-AO modules A and B are highlighted in blue. An inset indicating the layers of the specimen is shown on the right.